\documentclass[journal, lettersize]{IEEEtran}
\usepackage{amsmath,amssymb,amsfonts}
\allowdisplaybreaks
\usepackage{algorithm}
\usepackage{algpseudocode}
\usepackage{array}
\usepackage{tabularx,ragged2e,makecell}
\usepackage{seqsplit}

\usepackage[caption=false,font=footnotesize]{subfig}
\usepackage{textcomp}
\usepackage{stfloats}
\usepackage{url}
\usepackage{verbatim} 
\usepackage{graphicx} 
\usepackage{adjustbox}
\usepackage{balance}
\usepackage[normalem]{ulem}
\graphicspath{{figures/}{figures_revise/}{figures_latex/}}
\DeclareGraphicsExtensions{.pdf,.png,.jpg}
\usepackage{float}
\usepackage{tcolorbox}
\usepackage{xcolor}

\colorlet{blue}{black}
\colorlet{magenta}{black}

\tcbuselibrary{listings,skins,breakable}
\usepackage{tikz}
\usetikzlibrary{arrows.meta}
\usepackage{cite}
\usepackage[acronym]{glossaries}
\usepackage{booktabs}
\usepackage{supertabular}
\usepackage{placeins}
\usepackage{chngcntr}
\usepackage{hyperref}
\usepackage{cleveref}
\usepackage{standalone}
\usepackage{import}
\usepackage{msc}
\usepackage{listings}
\DeclareUnicodeCharacter{2248}{$\approx$}

\newcolumntype{Y}{>{\RaggedRight\arraybackslash}X}
\newcommand{\msg}[1]{\texttt{\seqsplit{#1}}}
\usepackage[utf8]{inputenc}
\usepackage{orcidlink}

\begin{document}
\newacronym{3gpp}{3GPP}{3rd Generation Partnership Project}
\newacronym{5gnr}{5GNR}{5G New Radio}

\newacronym{ack}{ACK}{Acknowledgement}
\newacronym{aiai}{AI-AI}{AI‑Native Air Interface}
\newacronym{ai}{AI}{Artificial Intelligence}
\newacronym{ai-core}{AI-Core}{Multi‑Agent Core Architecture}
\newacronym{agentgpt}{AgentGPT}{Domain‑Specific Large AI Model}
\newacronym{alibi}{ALiBi}{Attention with Linear Biases}
\newacronym{asn1}{ASN.1}{Abstract Syntax Notation}

\newacronym{bpe}{BPE}{Byte Pair Encoding}
\newacronym{bts}{BTS}{Base Transceiver Station}

\newacronym{cbpusch}{CB-PUSCH}{Contention‑Based Physical Uplink Shared Channel}
\newacronym{cp}{CP}{Control Plane}
\newacronym{cqi}{CQI}{Channel Quality Indicator}
\newacronym{csi}{CSI}{Channel State Information}
\newacronym{cu}{CU}{Central Unit}
\newacronym{cu-cp}{CU-CP}{Control Plane of Central Unit}
\newacronym{cu-up}{CU-UP}{User Plane of Central Unit}
\newacronym{dl}{DL}{Downlink}
\newacronym{drx}{DRX}{Discontinuous Reception}
\newacronym{du}{DU}{Distributed Unit}
\newacronym{o-du}{O-DU}{Open Distributed Unit}

\newacronym{e1}{E1}{Interface between CU-CP and CU-UP}
\newacronym{eni}{ENI}{Experiential Networked Intelligence}
\newacronym{earfcn}{EARFCN}{E-UTRA Absolute Radio Frequency Channel Number}
\newacronym{eutra}{E-UTRA}{Evolved UMTS Terrestrial Radio Access}

\newacronym{f1-c}{F1-C}{Control‑Plane Interface between CU‑CP and DU}
\newacronym{f1-u}{F1-U}{User‑Plane Interface between CU‑UP and DU}
\newacronym{f2-c}{F2-C}{Control‑Plane Interface between RRH and DU}
\newacronym{f2-u}{F2-U}{User‑Plane Interface between RRH and DU}
\newacronym{f1}{F1}{Interface between \gls{cu} and \gls{du}}
\newacronym{f2}{F2}{Interface between \gls{du} and \gls{ru}}

\newacronym{gemini}{Gemini}{Google DeepMind Multimodal LAM}
\newacronym{gptfour}{GPT-4o}{Generative Pre‑Trained Transformer 4o}
\newacronym{gnb}{gNB}{Next‑Generation Node B}
\newacronym{graph-rag}{Graph-RAG}{Graph Retrieval-Augmented Generation}

\newacronym{harq}{HARQ}{Hybrid Automatic Repeat Request}

\newacronym{ie}{IE}{Information Element}
\newacronym{int4}{INT4}{4-bit Integer Quantization}

\newacronym{l2}{L2}{Layer 2}
\newacronym{lam}{LAM}{Large AI Model}
\newacronym{lce}{LCE}{Learnable Control Embeddings}
\newacronym{llm}{LLM}{Large Language Model}
\newacronym{llama}{LLaMA}{Large Language Model Meta AI}
\newacronym{lora}{LoRA}{Low‑Rank Adaptation}
\newacronym{lte}{LTE}{Long Term Evolution}

\newacronym{mac}{MAC}{Medium Access Control}
\newacronym{ml}{ML}{Machine Learning}

\newacronym{netgpt}{NetGPT}{Telecom Knowledge Large AI Model}
\newacronym{nr}{NR}{New Radio}

\newacronym{pdb}{PDB}{Packet Delay Budget}
\newacronym{pdcp}{PDCP}{Packet Data Convergence Protocol}
\newacronym{pdu}{PDU}{Protocol Data Unit}
\newacronym{phy}{PHY}{Physical Layer}
\newacronym{ppo}{PPO}{Proximal Policy Optimization}
\newacronym{prb}{PRB}{Physical Resource Block}

\newacronym{qa}{QA}{Question-and-Answer}
\newacronym{qos}{QoS}{Quality of Service}

\newacronym{ran}{RAN}{Radio Access Network}
\newacronym{rb}{RB}{Resource Block}
\newacronym{rca}{RCA}{Root Cause Analysis}
\newacronym{rlc}{RLC}{Radio Link Control}
\newacronym{rrc}{RRC}{Radio Resource Control}
\newacronym{rrc-llm}{RRC-LLM}{Radio Resource Control–Integrated Large Language Model}
\newacronym{rrh}{RRH}{Remote Radio Head}
\newacronym{rf}{RF}{Radio Frequency}
\newacronym{ru}{RU}{Radio Unit}
\newacronym{rsrp}{RSRP}{Reference Signal Received Power}
\newacronym{rsrq}{RSRQ}{Reference Signal Received Quality}
\newacronym{rope}{RoPE}{Rotary Position Embeddings}

\newacronym{sbert}{SBERT}{Sentence BERT}
\newacronym{sdap}{SDAP}{Service Data Adaptation Protocol}
\newacronym{smc}{SMC}{State-Machine Conformance}
\newacronym{sora}{SORA}{OpenAI SORA Video Model}

\newacronym{tensorrt-llm}{TensorRT-LLM}{NVIDIA TensorRT-LLM}
\newacronym{transformerxl}{Transformer-XL}{Transformer-XL}
\newacronym{tti}{TTI}{Transmission Time Interval}

\newacronym{ue}{UE}{User Equipment}
\newacronym{ul}{UL}{Uplink}
\newacronym{up}{UP}{User Plane}
\newacronym{vllm}{vLLM}{vLLM Inference Engine}
\newacronym{xn}{Xn}{Inter‑gNB Interface}

\title{LLM-Based Emulation of the Radio Resource Control Layer: Toward AI-Native RAN Protocols}

\author{
  Ziming~Liu\orcidlink{0009-0002-6772-0005},~\IEEEmembership{Member,~IEEE},
  Bryan~Liu\orcidlink{0000-0001-7153-8885},~\IEEEmembership{Member,~IEEE},
  Alvaro~Valcarce\orcidlink{0000-0003-0400-3228},~\IEEEmembership{Senior~Member,~IEEE},
  and~Xiaoli~Chu\orcidlink{0000-0003-1863-6149},~\IEEEmembership{Senior~Member,~IEEE}
  \thanks{Z.~Liu and X.~Chu are with the School of Electrical and Electronic Engineering, The University of Sheffield, Sheffield S10 2TN, U.K. (e-mail: \{ziming.liu,x.chu\}@sheffield.ac.uk).}
  \thanks{B.~Liu and Á.~Valcarce are with Nokia Bell Labs, 12 Rue Jean Bart, 91300 Massy, France (e-mail: bryan.liu@unswalumni.com, alvaro.valcarce\_rial@nokia-bell-labs.com).}
  \thanks{This work was supported in part by the UK EPSRC grant EP/X038971/1 and the Horizon Europe Research and Innovation Program under grant 101086219.}
  \thanks{
  The authors acknowledge the use of resources provided by the Isambard-AI National AI Research Resource (AIRR). Isambard-AI is operated by the University of Bristol and is funded by the UK Government’s Department for Science, Innovation and Technology (DSIT) via UK Research and Innovation; and the Science and Technology Facilities Council [ST/AIRR/I-A-I/1023].
  McIntosh-Smith, S., Alam S. R. and Woods, C. (2024). "Isambard-AI: a leadership class supercomputer optimised specifically for Artificial Intelligence". https://doi.org/10.48550/arXiv.2410.11199
  }
  
  \thanks{Author accepted manuscript. Published in \emph{IEEE Journal on Selected Areas in Communications}, vol.~44, pp.~4319--4332, 2026. DOI: \href{https://doi.org/10.1109/JSAC.2026.3673712}{10.1109/JSAC.2026.3673712}. \copyright~2026 IEEE. Personal use of this material is permitted. Permission from IEEE must be obtained for all other uses, in any current or future media, including reprinting/republishing this material for advertising or promotional purposes, creating new collective works, for resale or redistribution to servers or lists, or reuse of any copyrighted component of this work in other works.}

}

\maketitle 
\begin{abstract}
\textcolor{black}{Integrating \glspl{lam} into 6G mobile networks is a key enabler of the \gls{aiai}, where protocol intelligence must scale beyond handcrafted logic. This paper presents, to our knowledge, the first standards-compliant emulation of the \gls{rrc} layer using a decoder-only \gls{lam} (\textsc{LLaMA}-class) fine-tuned with \gls{lora} on a multi-vendor corpus of real-world traces spanning both 5G and 4G systems. We treat \gls{rrc} as a domain-specific language and construct a segmentation-safe \gls{qa} dataset that preserves \gls{asn1} structure through linearization prior to \gls{bpe} tokenization. The proposed approach combines parameter-efficient adaptation with schema-bounded prompting to ensure syntactic and procedural fidelity. Evaluation introduces a standards-aware triad—\gls{asn1} conformance, field-level coverage analysis, and uplink-to-downlink state-machine checks—alongside semantic similarity and latency profiling across 120 configurations. On 30k 5G request–response pairs plus an additional 4.8k \gls{qa} turns from 4G sessions, our 8B model achieves a median cosine similarity of 0.97, a 61\% relative gain over a zero-shot baseline, while sustaining high conformance rates. These results demonstrate that \glspl{lam}, when augmented with protocol-aware reasoning, can directly orchestrate control-plane procedures, laying the foundation for the future \gls{ai}-native \gls{ran}.}
\end{abstract}
\glsresetall  

\section{Introduction}

\glspl{lam} are poised to transform wireless networks by enabling autonomous, cognitive capabilities that go beyond the scope of traditional optimization methods\cite{gao2025enabling}.
The forthcoming sixth generation (6G) of mobile networks is expected to incorporate \gls{ai} methods not merely as auxiliary optimisation tools but also as integral design primitives of the \gls{aiai}.
Recent proposals for an \gls{aiai} posit that \gls{ml} models could co‑design \gls{phy} and \gls{mac} procedures \cite{pl_1}, thereby enabling radio stacks that adapt automatically to dynamic service requirements, resource constraints, and propagation conditions \cite{Hoydis2021AI}.
The rapid evolution of \glspl{lam} \cite{OpenAI2025GPT4o, Google2023Gemini} strengthens this vision—when endowed with multi-modal perception and reasoning, such models become plausible agents for protocol‑centric tasks that demand both linguistic competence and structured decision making.
This raises a timely research challenge: determining the extent to which \glspl{lam} can directly engage in the generation and interpretation of control signaling in the \gls{ran}.

\textcolor{magenta}{This paper studies whether a decoder-only \gls{lam} can be fine-tuned to parse and generate 3GPP-compliant \gls{rrc} messages~\cite{3gpp38331}. Our contributions are threefold: (i) we curate multi-operator 4G/5G \gls{rrc} trace corpora and formulate an ASN.1-preserving \gls{qa} representation; (ii) we adapt the model via supervised fine-tuning enhanced with \gls{lora}~\cite{Hu2021LoRA} to emulate stateful \gls{rrc} procedures; and (iii) we quantify fidelity using standards-aware metrics—ASN.1 syntax validation~\cite{ITUT_ASN1_Intro}, schema-level field coverage, and UL$\rightarrow$DL state-machine conformance—alongside semantic similarity and latency. We provide that LAM-scale pretraining supplies compositional priors that generalize across 3GPP releases and vendor variants enabling multi-procedure reasoning even under incomplete context.}

\textcolor{magenta}{The 5G \gls{rrc} layer} orchestrates connection establishment, mobility and configuration between \gls{ue} and the \gls{gnb}.
Conceptually, \gls{rrc} messages form a specialized, domain‑specific language; learning this language is analogous to learning syntax and semantics in natural‑language processing.
Treating \gls{rrc} messages as such positions \gls{lam}‑based sequence models as natural emulators of the RRC layer.
Demonstrating accurate \gls{rrc} emulation would constitute an early, concrete instantiation of \gls{aiai} principles, with direct implications for the design of the 6G control plane.

Can the protocol literacy of a \gls{lam}—specifically, its ability to parse and generate 3GPP-compliant \gls{rrc} messages—be systematically improved, and through which fine‑tuning strategies?
This paper answers this question through a novel two‑phase methodology that (i) imparts \gls{rrc} knowledge to a decoder‑only \gls{lam} via supervised fine‑tuning enhanced with \gls{lora}, and (ii) evaluates the resulting model’s fidelity in emulating \gls{rrc} procedures under realistic network scenarios.
The findings offer evidence \textcolor{black}{about the feasibility of \gls{lam}‑based standards-compliant \gls{rrc} emulation}.

Without loss of generality, this work focuses on \emph{decoder‑only} transformer architectures (e.g., Llama‑class models \cite{Touvron2023LLaMA}).
Autoregressive decoding matches the turn‑taking structure of \gls{rrc} exchanges and permits key–value caching, which curtails inference latency and memory footprint, both critical for deployment at the \gls{gnb}.
Decoder‑only models thereby strike a favourable balance between capacity and computational efficiency, especially when combined with parameter‑efficient fine‑tuning techniques such as \gls{lora} \cite{Hu2021LoRA}.
\gls{lora} freezes the original model weights and introduces low‑rank update matrices, reducing trainable parameters by orders of magnitude while maintaining expressiveness; the approach mitigates both the cost of domain adaptation and the runtime overhead of inference, two primary barriers to operational use.

Training \glspl{lam} for specialist telecom domains demands curated datasets that capture the combinatorial breadth of protocol states, while deployment must meet stringent latency budgets to avoid impairing control‑plane responsiveness.
\textcolor{black}{To address these issues, we rebuilt two data corpuses of RRC traces (4G and 5G) with multi-operator coverage.
Then, we adopt a standards-aware evaluation suite that combines ASN.1~\cite{ITUT_ASN1_Intro} syntax validation, schema-level field coverage, and UL→DL state-machine conformance, reporting these jointly with semantic similarity and latency across 120 configurations.
Ultimately, we also outline a forward-looking engineering plan toward sub-100 ms inference latency via quantization~\cite{QLORA}, constrained decoding, KV caching and multi‑token draft‑verify decoding.}

\section{System Model}

This section delineates the architectural and operational principles underlying the proposed \gls{lam}-integrated \gls{gnb}.
We first describe the system architecture, detailing the functional decomposition of the disaggregated gNB and the embedding of a \gls{lam} within the RRC layer, while ensuring adherence to standardized interface definitions.
Then, the model-driven control-plane workflow is examined, encompassing the reception and contextual interpretation of uplink RRC messages and the autoregressive synthesis of standards-compliant downlink responses.

\begin{figure}[t!]
    \centering
    \includegraphics[width=1\linewidth]{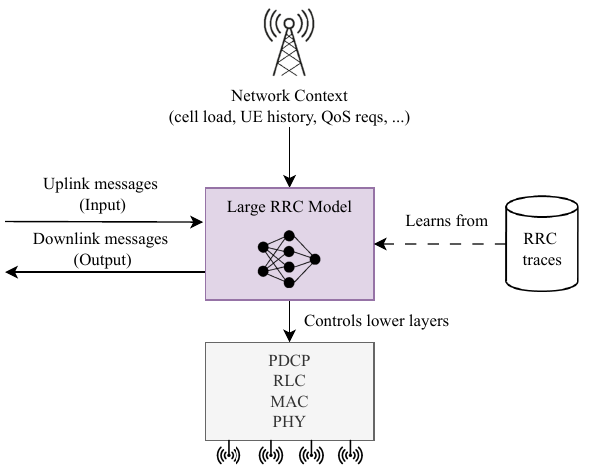}
    \caption{High-level concept of an AI-native gNB-side RRC layer powered by a \gls{llm}, illustrating its core inputs, outputs, learning source, and interaction with the protocol stack.}
    \label{fig:LLM-gNB-architecture-high}
\end{figure}

\subsection{Architecture and Functional Integration}

\begin{figure}[t!]
    \centering
    \includegraphics[width=1\linewidth]{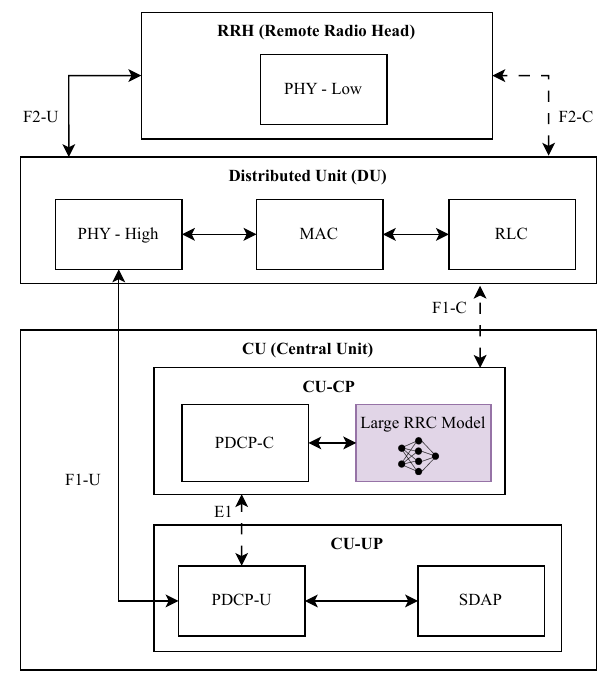}
    \caption{Low‑level architecture of an NR gNB disaggregating an LLM‑based RRC layer.}
    \label{fig:LLM-gNB-architecture-low}
\end{figure}

The proposed Large RRC Model integrates a decoder-only transformer-based \gls{lam} into the \gls{rrc} layer of a disaggregated \gls{gnb}, as depicted in Fig.~\ref{fig:LLM-gNB-architecture-high}.
Following the \gls{3gpp} NG-RAN standard, the \gls{gnb} is partitioned into a \gls{cu} and a \gls{du}, with the \gls{cu} further subdivided into separate \gls{cu-cp} and \gls{cu-up} entities. The \gls{cu-cp} hosts the \gls{rrc} layer, responsible for essential control-plane tasks such as connection establishment, radio bearer management, and mobility decisions.
Replacing the traditional rule-based logic, the embedded \gls{rrc} LLM serves as an intelligent decision engine within the \gls{rrc}, producing control messages and associated protocol decisions via autoregressive inference.
Conversely, the \gls{cu-up} encompasses user-plane layers, primarily \gls{sdap} and user-plane \gls{pdcp}, that manage the transport of user data. The \gls{du}, hosting lower-layer (e.g., \gls{rlc}, \gls{mac}, and PHY layers) protocols, interfaces directly with the \glspl{rrh} performing \gls{rf} and low-level physical-layer tasks.
Fig.~\ref{fig:LLM-gNB-architecture-low} illustrates all these interfaces and functions.
Integrating an \gls{llm} at the \gls{rrc} level is intended to preserve the external protocol behavior, leaving northbound and lateral interfaces to peer nodes unaltered, thus ensuring interoperability with existing network deployments.

As in standard 5G \gls{nr}, uplink \gls{rrc} messages from \glspl{ue}, such as Connection Requests or Measurement Reports, arrive at the \gls{cu-cp} via \gls{f1-c}, where they are processed by the embedded \gls{rrc} LLM, replacing the conventional RRC layer implementation.
The model interprets these inputs contextually, considering ongoing connection states, historical interactions, and dynamic network conditions to generate suitable downlink \gls{rrc} responses (e.g., Connection Setup or Mobility Reconfiguration). 
The generated messages maintain compliance with standardized \gls{3gpp} encoding, thus transparently replacing the message generation logic.
Downlink responses are then transmitted back to the \gls{ue} through the \gls{du} via the same interface.

\subsection{LLM-Driven Signaling Workflow and Model Operation}
Interactions between the \gls{rrc} layer and the user-plane modules are similarly orchestrated.
When the Large RRC Model decides to initiate or modify data radio bearers—potentially based on inferred context or changing service demands—it signals these decisions to the \gls{cu-up} over the \gls{e1}.
The \gls{cu-up} subsequently configures the corresponding \gls{sdap} and \gls{pdcp} entities, enabling adaptive management of the user-plane traffic.
Meanwhile, data-plane traffic flows remain unmodified and proceed independently over the established user-plane interfaces (\gls{f1-u}, \gls{f2-u}).

\begin{figure*}[ht]
    \centering
    \includegraphics[width=1\linewidth]{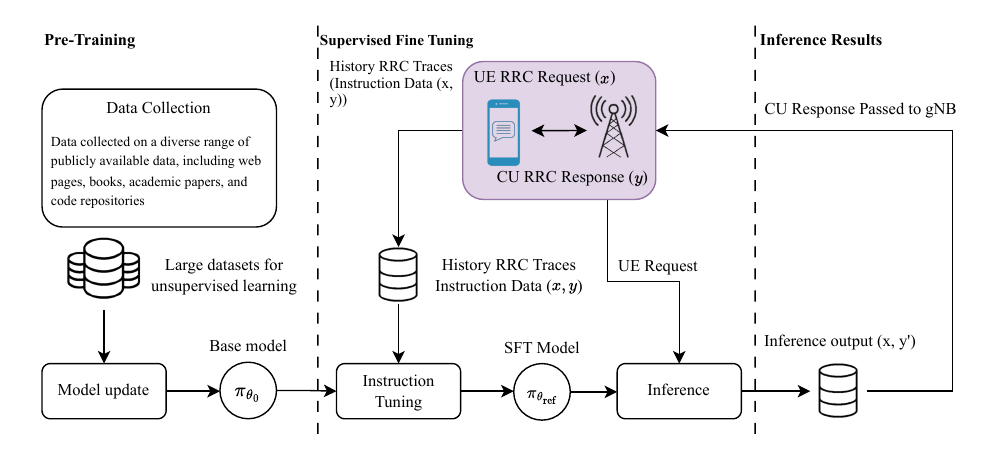}
    \caption{Data flow for fine-tuning and inference in the LLM-based RRC system.}
    \label{fig:LLM-training}
\end{figure*}
Fig.~\ref{fig:LLM-training} illustrates the fine-tuning and inference workflow specific to the proposed \gls{rrc} \gls{llm}.
The decoder-only transformer is taught \gls{rrc} domain knowledge using supervised fine-tuning on a curated dataset of historical \gls{rrc} message traces collected from a real-world 5G network.
This specialized fine-tuning imparts explicit protocol knowledge to the model, enabling it to emulate standard \gls{rrc} behaviors accurately during live inference.
Operationally, the trained model functions continuously within the \gls{cu-cp}, dynamically generating control-plane messages as new \gls{ue} requests are received.

\section{Fine-tuning of LLM for RRC Message Processing}
\textcolor{blue}{This section describes the pipeline for adapting a decoder-only \gls{llm} to emulate \gls{rrc} procedures. Our approach consists of three components: (i) dataset construction, where raw traces are reorganized into segmentation-safe uplink–downlink \gls{qa} pairs; (ii) parameter-efficient supervised fine-tuning using \gls{lora}; and (iii) latency-aware evaluation to assess feasibility under control-plane timing constraints. Earlier traces often exceeded context limits, requiring ad-hoc segmentation that broke procedural continuity. Our design eliminates this by enforcing per-procedure boundaries.}

\begingroup %
\color{black}
\subsection{Datasets: Construction, Scope, and Coverage}
\label{sec:DatasetConstruction}

We use two complementary corpora with a common, segmentation-safe construction pipeline:
\begin{itemize}
  \item \textbf{NR dataset (proprietary, single-operator test setups):}
  \(\sim\)30{,}247 UL\(\leftrightarrow\)DL RRC pairs extracted from 2{,}524 sessions collected in controlled test environments.
  \item \textbf{LTE dataset (proprietary road tests across two networks):}
  \(\sim\)4{,}808 QA turns across 1{,}601 sessions collected via real-world drive tests.
\end{itemize}

\paragraph*{Segmentation-safe UL\(\rightarrow\)DL QA construction}
Raw RRC traces are sessionized and reorganized into UL--DL QA pairs preserving procedural causality and ASN.1 structure (cf.~\Cref{fig:rrc_trace}). Each session is capped in turns to respect context limits, with adjacent UL messages concatenated when appropriate (e.g., \texttt{rrcReconfigurationComplete} \(+\) \texttt{measurementReport}). The resulting per-turn inputs (\(X\)) and targets (\(Y\)) follow the mapping in \cref{subsec:math_formulation} and Eq.~\ref{eq:conditional}--\ref{eq:inference}.

\paragraph*{Schema-bounded prompting and radio constraints}
To enforce structural validity, we inject a session-specific, size-bounded ASN.1 micro-schema (see Fig.~\ref{fig:lte-template-corr} in the appendix) into the system prompt together with radio constraints (e.g., allowed EARFCNs). This preserves field names/order and prevents extraneous elements without token-level grammar overhead. This micro-schema mechanism was applied only to the LTE corpora.

\paragraph*{Attribution and availability}
Both corpora are \emph{private and proprietary}. Public LTE datasets/simulators exist (e.g., \cite{liu_rrc_2025,liu_nrRRC_simulator_2025}), but are synthetic and were \emph{not used} in this work.

\paragraph*{Coverage}
Table~\ref{tab:rrc-breakdown} summarizes the observed message-type counts across the two corpora. Major RRC procedures (connection establishment, initial security activation, reconfiguration, release, capability transfer) are well represented in both RATs.
\endgroup

\begin{table}[t]
  \centering
  \caption{Breakdown of LTE and NR RRC message types observed in the proprietary field logs. A \textemdash{} denotes not observed for that RAT.}
  \label{tab:rrc-breakdown}
  \label{tab:rrc-breakdown-lte}
  \label{tab:rrc-breakdown-nr}
  \begin{tabular*}{\linewidth}{@{}l@{\extracolsep{\fill}}rr@{}}
    \toprule
    Message & LTE Count & NR Count \\
    \midrule
    \texttt{dlInformationTransfer} & 149 & 1\,743 \\
    \texttt{measurementReport} & 229 & 43\,410 \\
    \texttt{mobilityFromNRCommand} & \textemdash{} & 92 \\
    \texttt{paging} & 1\,016 & \textemdash{} \\
    \texttt{rrcConnectionReconfiguration} & 701 & \textemdash{} \\
    \texttt{rrcConnectionReconfigurationComplete} & 891 & \textemdash{} \\
    \texttt{rrcConnectionRelease} & 110 & \textemdash{} \\
    \texttt{rrcConnectionRequest} & 662 & \textemdash{} \\
    \texttt{rrcConnectionReestablishment} & 3 & \textemdash{} \\
    \texttt{rrcConnectionReestablishmentRequest} & 5 & \textemdash{} \\
    \texttt{rrcConnectionSetup} & 775 & \textemdash{} \\
    \texttt{rrcConnectionSetupComplete} & 879 & \textemdash{} \\
    \texttt{rrcReconfiguration} & \textemdash{} & 16\,034 \\
    \texttt{rrcReconfigurationComplete} & \textemdash{} & 16\,099 \\
    \texttt{rrcRelease} & \textemdash{} & 4\,267 \\
    \texttt{rrcReestablishment} & \textemdash{} & 28 \\
    \texttt{rrcReestablishmentComplete} & \textemdash{} & 28 \\
    \texttt{rrcReestablishmentRequest} & \textemdash{} & 119 \\
    \texttt{rrcSetup} & \textemdash{} & 4\,072 \\
    \texttt{rrcSetupComplete} & \textemdash{} & 4\,068 \\
    \texttt{rrcSetupRequest} & \textemdash{} & 4\,009 \\
    \texttt{securityModeCommand} & 653 & 3\,827 \\
    \texttt{securityModeComplete} & 693 & 3\,824 \\
    \texttt{systemInformation} & 95 & \textemdash{} \\
    \texttt{systemInformationBlockType1} & 265 & \textemdash{} \\
    \texttt{ueCapabilityEnquiry} & 633 & 876 \\
    \texttt{ueCapabilityInformation} & 6 & 873 \\
    \texttt{ueInformationRequest-r9} & 3 & \textemdash{} \\
    \texttt{ueInformationResponse-r9} & 134 & \textemdash{} \\
    \texttt{ulInformationTransfer} & 55 & 1\,924 \\
    \midrule
    \textbf{Total} & \textbf{7\,957} & \textbf{105\,293} \\
    \bottomrule
  \end{tabular*}
\end{table}

\begin{figure}[t]
    \centering
    \includestandalone[width=\columnwidth]{figures_latex/rrc_trace}
    \caption{Example of a pre-processed \gls{rrc} trace segment. Messages are structured into Question (Q) / Answer (A) pairs for LLM training, where 'Q' denotes the input RRC message(s) (X) and 'A' the target response (Y). Note the consolidation of multiple physical messages (e.g., \emph{rrcReconfigurationComplete} and \emph{measurementReport}) into a single logical request, labeled here as Q4(concat) and Q6(concat).}
    \label{fig:rrc_trace}
\end{figure}

%

%

%

%
%
%
%
%
%
%

%

%
%
%
%

%
%

%
%

%
%
%
%

%
%

%
%
%
%
%
%
%
%
%
%
%

%
%
%
%
%
%
%
%
%
%
%

%
%
%

%
%

%
%

%
%
%
%
%
%
%
%
%

%
%

%

%
%

%

%

%
%

%

%

%
%
%
%
%
%
%
%
%
%
%
%
%
%
%
%
%
%
%
%
%
%
%
%
%

%

%

%

%
%
%
%
%
%
%
%
%
%
%
%
%
%
%
%
%
%
%
%
%

%
%
%

%
%
%
%
%
%
%

%
%
%

%
%

%
%

%
%
%
%
%
%
%
%

%

\subsection{\textcolor{blue}{Supervised Fine-Tuning Framework}}
\label{subsec:math_formulation}

\textcolor{blue}{We model \gls{rrc} request–response exchanges as a conditional generation task over uplink–downlink pairs. Let $X$ denote the uplink message sequence and $Y$ the corresponding downlink response. The objective is to estimate the conditional probability:}
\begin{equation}
\pi_\theta(Y \mid X) = \prod_{t=1}^{T} P(y_t \mid y_{<t}, X; \theta),
\label{eq:conditional}
\end{equation}
\textcolor{blue}{where $\theta$ are trainable parameters, $y_t$ is the token generated at step $t$, and $T$ is the sequence length. This formulation aligns with the \gls{qa}-style mapping introduced in Sec.~\ref{sec:DatasetConstruction}.}
\textcolor{magenta}{$X$ and $Y$ are tokenized using the same \gls{bpe} tokenizer as the Llama backbone.
Therefore, $T$ denotes the number of \gls{bpe} tokens in the response sequence.}

\textcolor{blue}{To adapt a decoder-only \gls{llm} efficiently, we apply \gls{lora}, which introduces low-rank updates to frozen backbone weights. For a weight matrix $W_0 \in \mathbb{R}^{d \times k}$, the update is:}
\begin{equation}
W_\theta = W_0 + AB,
\label{eq:lora}
\end{equation}
\textcolor{blue}{where $A \in \mathbb{R}^{d \times r}$ and $B \in \mathbb{R}^{r \times k}$ are trainable factors with rank $r \ll \min\{d,k\}$. This reduces trainable parameters by orders of magnitude while preserving expressiveness. The fine-tuning objective minimizes the negative log-likelihood:}
\begin{equation}
\mathcal{L}_{\text{SFT}}(\theta) = -\frac{1}{N}\sum_{i=1}^{N}\log \pi_\theta(Y^{(i)} \mid X^{(i)}).
\label{eq:sft}
\end{equation}
\textcolor{blue}{Here, $N$ denotes the total number of aligned uplink–downlink pairs in the supervised training set. Inference uses greedy decoding under the fine-tuned model:}
\begin{equation}
\hat{Y} = \arg\max_{Y} \pi_\theta(Y \mid X).
\label{eq:inference}
\end{equation}

\begin{algorithm}[h]
\caption{RRC-LLM Training and Inference Pipeline}
\label{alg:rrc_pipeline}
\begin{algorithmic}[1]
\Require Raw RRC traces $\mathcal{T}_{\text{raw}}$, pre-trained model $\pi_{\theta_{\text{init}}}$
\Ensure Fine-tuned model $\pi_{\theta^*}$

\State \textbf{Phase 1: Dataset Construction}
\State $\mathcal{D} \gets \texttt{Preprocess}(\mathcal{T}_{\text{raw}})$
\Comment{Filter, align, and extract $(X, Y)$ pairs}
\ForAll{$(X^{(i)}, Y^{(i)}) \in \mathcal{D}$}
    \State $X^{(i)} \gets \texttt{BPE\_Tokenize}(X^{(i)})$
    \State $Y^{(i)} \gets \texttt{BPE\_Tokenize}(Y^{(i)})$
\EndFor
\State \textbf{Phase 2: Supervised Fine-Tuning (SFT)}
\State $\pi_{\theta_{\text{ref}}} \gets \texttt{LoRA\_Init}(\pi_{\theta_{\text{init}}})$
\For{$e = 1$ to $E_{\text{SFT}}$}
    \ForAll{$(X^{(i)}, Y^{(i)}) \in \mathcal{D}$}
        \State $\mathcal{L}_{\text{SFT}} \gets -\log \pi_{\theta_{\text{ref}}}(Y^{(i)} \mid X^{(i)})$
        \State Update $\theta_{\text{ref}}$ using Adam on $\mathcal{L}_{\text{SFT}}$
    \EndFor
\EndFor
\State \textbf{Phase 3: Inference}
\State Given new uplink message $X$, compute
\State \hspace{2em} $\hat{Y} \gets \arg\max_Y \pi_{\theta}(Y \mid X)$
\Return $\hat{Y}$
\end{algorithmic}
\end{algorithm}

\subsection{Training Configuration and Convergence Behavior}

\begin{figure}[t]
    \centering
    \includegraphics[width=1\linewidth]{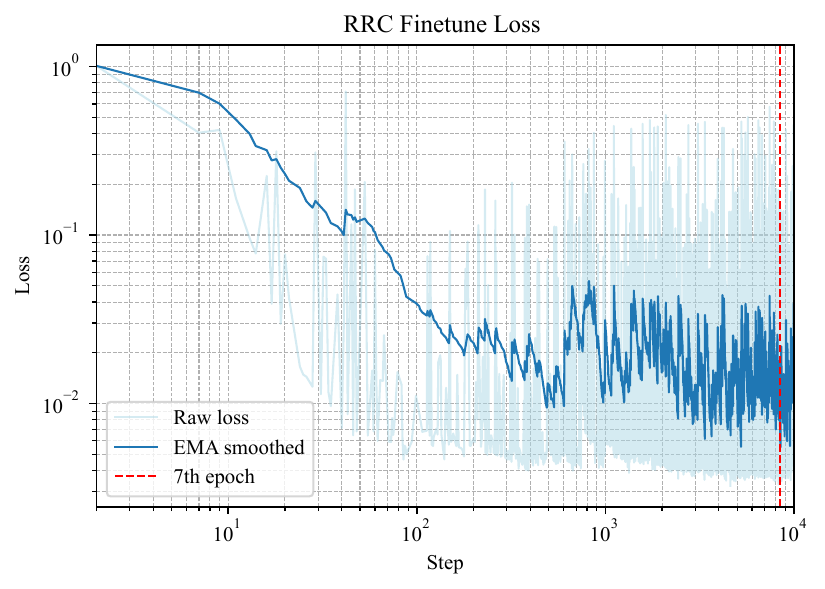}
    \caption{Training loss trajectory on the \textbf{NR dataset} (proprietary corpus). The pale blue trace is the raw loss, and the dark blue curve is its exponential moving average (EMA). The vertical red dashed line marks the epoch selected for downstream evaluation.
    }
    \label{fig:loss_curve_checkpoint}
\end{figure}

\textcolor{blue}{Supervised fine-tuning of the Llama-3 8B backbone was performed under the configuration in Table~\ref{tab:training_config}. The loss decreased by two orders of magnitude within the first $10^3$ steps and stabilized after $\sim$8.4k steps, as shown in Fig.~\ref{fig:loss_curve_checkpoint}. The EMA-smoothed trace flattens near $10^{-2}$, indicating convergence. We select the checkpoint at this point to capture $>\!95\%$ of attainable loss reduction while avoiding overfitting and unnecessary GPU hours. Fig.~\ref{fig:models_methods_grid} further compares training and validation loss across backbones and LoRA ranks, highlighting stability trends and the effect of parameter‑efficient tuning.
}

\textcolor{blue}{This trajectory exhibits three regimes: (i) rapid descent during warm-up ($0$--$10^3$ steps), (ii) gradual improvement ($10^3$--$8.4\times10^3$), and (iii) saturation beyond $8.4\times10^3$. The EMA curve mitigates variance from the small effective batch size ($2\times4096$ tokens). Validation on a disjoint trace suite confirmed generalization without requiring full validation passes during training. Fig.~\ref{fig:loss_curve_checkpoint} illustrate the convergence profile. The selected checkpoint (red dashed line) balances convergence speed and stability, ensuring fidelity to RRC message structures while minimizing compute overhead.}

\textcolor{magenta}{We note that earlier iterations of this training pipeline
experienced frequent interruptions (e.g., SSD stalls, NCCL watchdog timeouts). These were identified as infrastructure-specific bottlenecks (insufficient storage I/O and driver-level software bugs) rather than model instability. Following the migration to the updated cluster (\Cref{tab:server_specs}), the training process proved robust, achieving convergence without any hardware-induced interruptions.}

\begin{table}%
\caption{Supervised fine-tuning hyperparameters}
\label{tab:training_config}
\centering
\begin{tabular}{lcc}
\toprule
\textbf{Parameter} & \textbf{NR Dataset} & \textbf{LTE Dataset} \\
\midrule
Model & Llama-3 8B & Llama-3.x (1B–8B) \\
Max Sequence Length & 4096 & 4096 \\
Precision & bf16 & bf16 \\
Optimizer & AdamW & AdamW \\
Learning Rate & $2\times10^{-5}$ & \makecell{$2\times10^{-5}$ (full FT)\\$3\times10^{-4}$ (LoRA)} \\
Loss Function & CrossEntropy & CrossEntropy \\
LR schedule & -- & \makecell{Cosine decay\\+ 100-step warmup} \\
LoRA Config & -- & $\alpha=2r$, dropout=0 \\
\bottomrule
\end{tabular}
\end{table}

\subsection{\textcolor{blue}{Decoding regimes}}
\textcolor{blue}{We evaluate three decoding strategies that govern prompt structure and output constraints:}

\begin{itemize}
\item \textcolor{blue}{\textbf{NoSys}: A generic system prompt with minimal guidance, allowing unconstrained generation. This regime prioritizes raw language-model priors and serves as a lower-bound baseline.}
\item \textcolor{blue}{\textbf{RRC}: A domain-oriented prompt that specifies the \gls{rrc} context and message type but does not enforce strict schema constraints. This improves semantic alignment but leaves structural validity largely unchecked.}
\item \textcolor{blue}{\textbf{RRC\_constrain}: A schema-bounded prompt that injects a size-limited \gls{asn1} micro-schema and radio constraints (e.g., allowed \gls{earfcn}) into the system message. Generation is restricted to fields declared in this schema, preserving field order and prohibiting extraneous elements. This regime is designed to maximize standards compliance without incurring token-level grammar overhead.}
\end{itemize}

\textcolor{blue}{Unless otherwise stated, all reported results in Sec.~\ref{sec:performance_evaluation} use \texttt{RRC\_constrain} for tuned configurations, as it delivers near-ceiling \gls{asn1}/\gls{smc} pass rates while maintaining practical latency.}
\textcolor{magenta}{Crucially, this schema-bounded approach mitigates the structural violations observed in unconstrained baselines (e.g., the duplication of fields). By explicitly injecting the session-specific ASN.1 definitions into the context, we leverage the model's in-context learning capabilities to ground the generation process. This ensures that the model attends strictly to the fields declared in the active micro-schema, effectively preventing the hallucination of extraneous or repetitive elements without requiring computationally expensive token-level grammar constraints.}

{\color{blue}
\subsection{LTE Corpus: Multi-Backbone and LoRA Rank Ablations}

\textcolor{blue}{To evaluate cross-RAT robustness, we repeated the fine-tuning pipeline on an LTE-specific corpus (1,601 sessions, 4,808 QA turns) using the same segmentation-safe design as in Sec.~\ref{sec:DatasetConstruction}. This LTE dataset captures connection setup, security activation, and reconfiguration procedures under diverse mobility scenarios.}

\textcolor{blue}{We benchmarked four Llama-3 backbones (8B, 8B-3.1, 3B, 1B) under full fine-tuning and LoRA with ranks $r\in\{4,8,16\}$. All runs used bf16 precision, AdamW optimizer, and token-level cross-entropy loss. LoRA adapters were applied to attention projections and MLP layers with scaling $\alpha=2r$ and dropout $=0$. Learning rates followed standard practice: $2\times10^{-5}$ for full FT and $3\times10^{-4}$ (cosine decay, 100-step warmup) for LoRA.~\ref{fig:loss_curve_checkpoint}}

\textcolor{blue}{Table~\ref{tab:lora-ablation} summarizes the best validation loss per configuration. On 8B backbones, higher-rank LoRA ($r=16$) matches or slightly surpasses full fine-tuning, while on 3B/1B models, updating all weights remains advantageous. Across all settings, the gap between the best two methods is small ($\Delta\approx6\times10^{-4}$), confirming that parameter-efficient adaptation can approach full FT performance on large models.}

\begin{table}[!t]
\centering
\caption{\textcolor{blue}{LTE Corpus: Validation Loss at Best Checkpoint (Lower is Better)}}
\label{tab:lora-ablation}
\begin{tabular}{lcccc}
\hline
\textbf{Backbone} & \textbf{Full FT} & \textbf{LoRA r4} & \textbf{LoRA r8} & \textbf{LoRA r16} \\
\hline
Llama-3 8B        & 0.02017 & 0.02038 & 0.02062 & \textbf{0.01984} \\
Llama-3.1 8B      & 0.02836 & 0.02131 & 0.02152 & \textbf{0.02096} \\
Llama-3.2 3B      & \textbf{0.01915} & 0.02125 & 0.02232 & 0.02207 \\
Llama-3.2 1B      & \textbf{0.02017} & 0.02130 & 0.02106 & 0.02105 \\
\hline
\end{tabular}
\end{table}

\textcolor{blue}{These results highlight two trends: (i) larger backbones benefit from higher LoRA ranks, reducing the need for full FT; (ii) smaller models favor full updates for best accuracy. This informs the latency–fidelity trade-offs explored in Sec.~\ref{sec:performance_evaluation}.}

\begin{table*}[t]
\centering
\caption{Curated results on the LTE dataset.}
\label{tab:new_benchmark_top18}
\begin{tabular}{l l l l r r r r r r r r}
\toprule
Backbone & Quant & Strategy & Tuning & \multicolumn{2}{c}{Latency (ms)} & \multicolumn{2}{c}{Schema} & \multicolumn{2}{c}{Similarity} & \multicolumn{2}{c}{Pass Rate} \\
\cmidrule(lr){5-6} \cmidrule(lr){7-8} \cmidrule(lr){9-10} \cmidrule(lr){11-12}
 &  &  &  & Med. & Avg. & Avg & Med & Avg & Med & ASN & SMC \\
\midrule
 L-3 8B & FP16 & RRC\_constrain & lora-r16 & 3544 & 5258.0 & 0.976 & 1.000 & 0.992 & 1.000 & 0.995 & 0.993 \\
 L-3 8B & Q4\_K\_M & RRC\_constrain & lora-r16 & 2556 & 3854.1 & 0.969 & 1.000 & 0.990 & 1.000 & 0.989 & 0.988 \\
 L-3 8B & FP16 & RRC & lora-r16 & 3418 & 4280.4 & 0.467 & 0.333 & 0.723 & 0.669 & 0.023 & 0.000 \\
 L-3 8B & Q4\_K\_M & RRC & lora-r16 & 2519 & 2529.7 & 0.439 & 0.333 & 0.726 & 0.666 & 0.003 & 0.002 \\
 L-3 8B & FP16 & NoSys & lora-r16 & 1987 & 2011.7 & 0.231 & 0.241 & 0.480 & 0.478 & 0.003 & 0.002 \\
 L-3 8B & Q4\_K\_M & NoSys & lora-r16 & 1431 & 1442.1 & 0.235 & 0.239 & 0.491 & 0.490 & 0.012 & 0.004 \\

 L-3.1 8B & FP16 & RRC\_constrain & lora-r16 & 3609 & 5367.3 & 0.969 & 1.000 & 0.991 & 1.000 & 0.995 & 0.994 \\
 L-3.1 8B & Q4\_K\_M & RRC\_constrain & lora-r16 & 2581 & 3939.0 & 0.964 & 0.987 & 0.989 & 1.000 & 0.904 & 0.902 \\
 L-3.1 8B & FP16 & RRC & lora-r16 & 3364 & 2716.1 & 0.523 & 0.675 & 0.643 & 0.650 & 0.001 & 0.000 \\
 L-3.1 8B & Q4\_K\_M & RRC & lora-r16 & 1379 & 1653.6 & 0.371 & 0.262 & 0.579 & 0.508 & 0.001 & 0.000 \\
 L-3.1 8B & FP16 & NoSys & lora-r16 & 2193 & 2243.0 & 0.211 & 0.224 & 0.477 & 0.478 & 0.003 & 0.002 \\
 L-3.1 8B & Q4\_K\_M & NoSys & lora-r16 & 1501 & 1537.8 & 0.213 & 0.220 & 0.485 & 0.482 & 0.001 & 0.000 \\

 L-3.2 3B & FP16 & RRC\_constrain & full & 2341 & 3501.9 & 0.969 & 1.000 & 0.991 & 1.000 & 0.993 & 0.991 \\
 L-3.2 3B & Q4\_K\_M & RRC\_constrain & full & 1733 & 2658.1 & 0.954 & 1.000 & 0.987 & 1.000 & 0.973 & 0.968 \\
 L-3.2 3B & FP16 & RRC & full & 2157 & 2939.1 & 0.408 & 0.353 & 0.688 & 0.700 & 0.000 & 0.000 \\
 L-3.2 3B & Q4\_K\_M & RRC & full & 2170 & 2920.6 & 0.440 & 0.400 & 0.689 & 0.698 & 0.001 & 0.000 \\
 L-3.2 3B & FP16 & NoSys & full & 1755 & 1848.0 & 0.214 & 0.223 & 0.466 & 0.467 & 0.011 & 0.004 \\
 L-3.2 3B & Q4\_K\_M & NoSys & full & 1367 & 1452.8 & 0.201 & 0.207 & 0.497 & 0.493 & 0.040 & 0.015 \\

 L-3.2 1B & FP16 & RRC\_constrain & full & 2093 & 3867.2 & 0.964 & 1.000 & 0.988 & 1.000 & 0.993 & 0.989 \\
 L-3.2 1B & Q4\_K\_M & RRC\_constrain & full & 2025 & 3585.8 & 0.934 & 0.987 & 0.981 & 1.000 & 0.888 & 0.883 \\
 L-3.2 1B & FP16 & RRC & full & 589 & 1251.0 & 0.336 & 0.333 & 0.652 & 0.649 & 0.000 & 0.000 \\
 L-3.2 1B & Q4\_K\_M & RRC & full & 428 & 1368.8 & 0.293 & 0.286 & 0.662 & 0.665 & 0.000 & 0.000 \\
 L-3.2 1B & FP16 & NoSys & full & 1142 & 1390.6 & 0.202 & 0.200 & 0.521 & 0.514 & 0.024 & 0.009 \\
 L-3.2 1B & Q4\_K\_M & NoSys & full & 1209 & 1335.5 & 0.189 & 0.182 & 0.501 & 0.502 & 0.024 & 0.016 \\
\bottomrule
\end{tabular}
\end{table*}

\begin{figure*}[t]
    \centering
    \includegraphics[width=\linewidth]{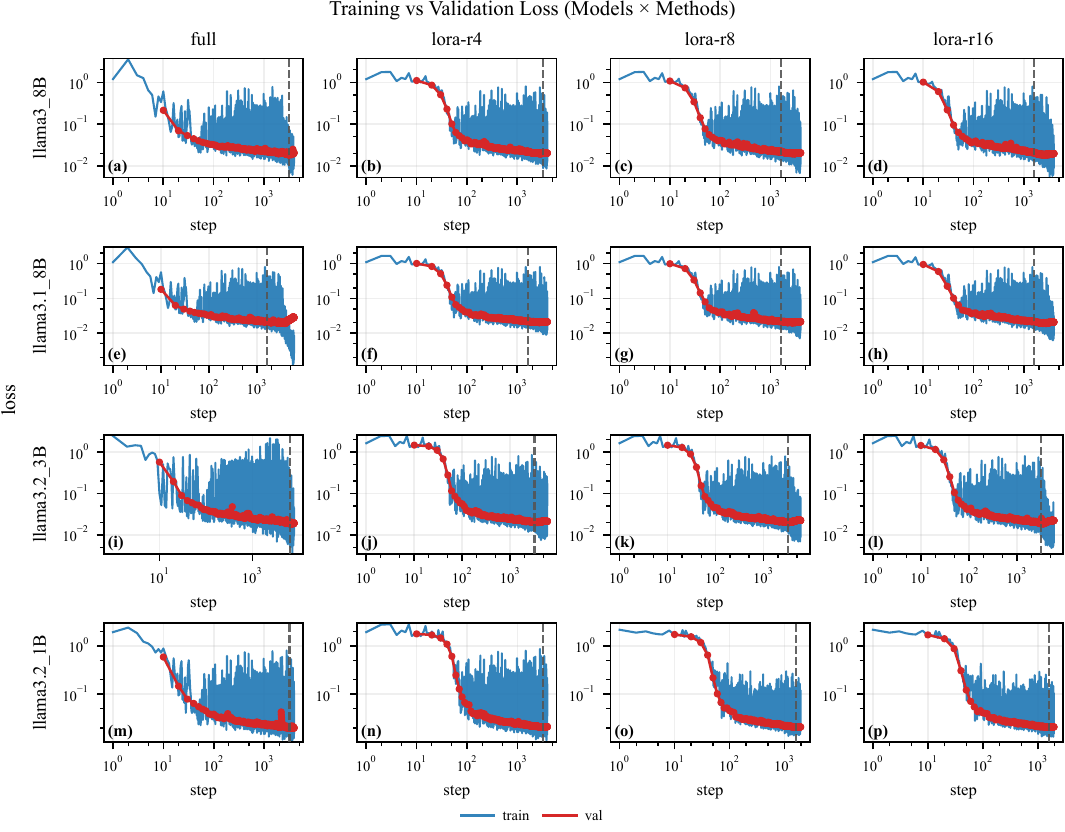}
    \caption{\textbf{Training vs.\ validation loss across models and methods on the \textbf{LTE dataset} (public corpus).}
Columns: full fine‑tuning and LoRA with $r\!\in\!\{4,8,16\}$. Rows: \textsc{Llama‑3}~8B, \textsc{Llama‑3.1}~8B, \textsc{Llama‑3.2}~3B, \textsc{Llama‑3.2}~1B. Blue traces denote training loss, red traces denote validation loss; the vertical dashed line indicates the checkpoint selected for evaluation.}
    \label{fig:models_methods_grid}
\end{figure*}

}
\section{Performance and Evaluation}
\label{sec:performance_evaluation}

\subsection{\textcolor{blue}{Evaluation metrics and protocol conformance}}
\textcolor{blue}{We evaluate the generated \gls{dl} \gls{rrc} messages along three independent and complementary axes: (i) semantic proximity to the reference using \gls{sbert}-based \cite{reimers-2019-sentence-bert} cosine similarity, (ii) \gls{asn1} syntax validity via encode–decode round-trip under a session-specific micro-schema, and (iii) \gls{ul}\,$\rightarrow$\,\gls{dl} \gls{smc}. These metrics are reported jointly with latency in Tables~\ref{tab:new_benchmark_top18}--\ref{tab:ablation_global_medians}.}

\begin{figure}[t]
  \centering
  \includegraphics[width=0.95\linewidth]{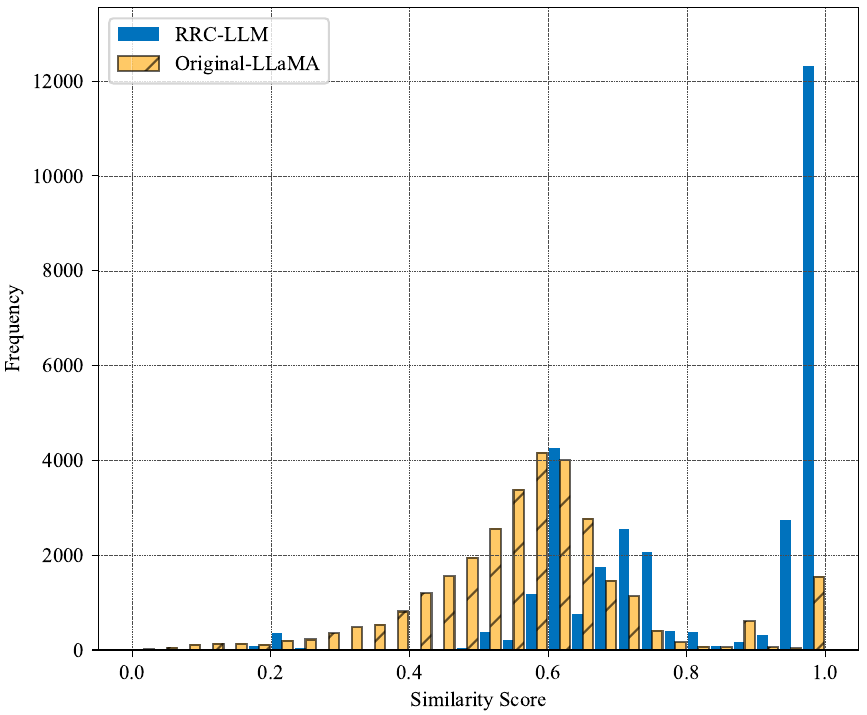}
  \caption{\textcolor{blue}{Distribution of SBERT-based cosine similarity scores on the NR dataset (proprietary corpus) using the LLaMA-3 8B backbone. The histogram contrasts the fine-tuned model against zero-shot baselines.}}
  \label{fig:similarity_distribution}
\end{figure}

\vspace{0.25em}
\paragraph{\textcolor{blue}{Semantic similarity (\gls{sbert} cosine)}}
\textcolor{blue}{For each test sample $i$, let $Y_i$ be the ground-truth \gls{dl} message and $\hat{Y}_i$ the model output given the \gls{ul} context $X_i$. We encode both with a frozen \gls{sbert} encoder $f_{\text{sbert}}(\cdot)$ and mean-pool to sentence embeddings $u_i$ and $v_i$:}
\begin{equation}
\textcolor{blue}{u_i=\frac{1}{|Y_i|}\sum_{t} f_{\text{sbert}}(Y_i)_t,\qquad
v_i=\frac{1}{|\hat{Y}_i|}\sum_{t} f_{\text{sbert}}(\hat{Y}_i)_t.}
\label{eq:pool}
\end{equation}
\textcolor{blue}{The cosine similarity is $s_i=\frac{u_i^\top v_i}{\|u_i\|\,\|v_i\|}\in[-1,1]$. We report $s_i$ directly. When combining semantic and structural metrics, a $[0,1]$ range is preferred, so we use $\tilde{s}_i=\frac{s_i+1}{2}$.
Fig.~\ref{fig:similarity_distribution} illustrates the distribution of SBERT-based similarity scores for the \textbf{5G~NR field corpus} under the \textbf{LLaMA-3~8B backbone} after fine-tuning with \gls{lora}.}

\vspace{0.25em}
\paragraph{\textcolor{blue}{\gls{asn1} round-trip validity and schema recall}}
\textcolor{blue}{For each session $i$, we construct a trimmed \gls{asn1} micro-schema $\mathcal{S}^{(i)}$ (\emph{messages and IEs relevant to that session}) and perform a strict encode--decode round-trip on $\hat{Y}_i$. The binary pass indicator is}
\begin{equation}
\textcolor{blue}{\mathrm{asn}_i \;=\; \mathbb{I}\Big[\mathrm{Dec}_{\mathcal{S}^{(i)}}\big(\mathrm{Enc}_{\mathcal{S}^{(i)}}(\hat{Y}_i)\big)\ \text{succeeds $\land$ plausible}\Big].}
\label{eq:asn}
\end{equation}
\textcolor{blue}{To measure field-level faithfulness beyond mere text similarity, we compare the set of normalized \gls{asn1} \gls{ie} names present in the reference and the hypothesis. Let $W(\cdot)$ extract this set; for non-empty $W(Y_i)$, the per-sample schema recall is}
\begin{equation}
\textcolor{blue}{r_i \;=\; \frac{\big|\,W(Y_i)\cap W(\hat{Y}_i)\,\big|}{\big|\,W(Y_i)\,\big|}\in[0,1].}
\label{eq:schema}
\end{equation}

\vspace{0.25em}
\paragraph{\textcolor{blue}{UL$\rightarrow$DL state-machine conformance (\gls{smc})}}
\textcolor{blue}{From the \gls{ul} input we derive the set $U_i$ of \gls{ul} message types; from $\hat{Y}_i$ we extract the \gls{dl} type $d_i$ and, when present, the RRC transaction identifier $\mathrm{TxId}(\hat{Y}_i)$. Let $\mathcal{A}(u)$ denote the set of allowed \gls{dl} types given a \gls{ul} type $u$ (\emph{strong edges}); for a set $U$, write $\mathcal{A}(U)=\bigcup_{u\in U}\mathcal{A}(u)$. The strict binary \gls{smc} pass is then}
\begin{equation}
\textcolor{blue}{
\mathrm{smc}_i = \mathbb{I}\!\Big[ d_i \in \mathcal{A}(U_i)\ \wedge\ \mathrm{TxId}(\hat{Y}_i)=\mathrm{TxId}(Y_i)\Big].
}
\label{eq:smc}
\end{equation}
\textcolor{blue}{Here, the transaction-ID equality is enforced only when both sides carry an ID. We additionally report a \emph{weak} profile in which broadcast/weak edges are admitted (not shown for brevity when identical).}

\vspace{0.25em}
\paragraph{\textcolor{blue}{Reporting}}
\textcolor{blue}{In Tables~\ref{tab:new_benchmark_top18}--\ref{tab:ablation_global_medians}, we summarize results by the distribution (median and dispersion) of $s_i$ (or $\tilde{s}_i$), the dataset-level \gls{asn1} pass rate $\frac{1}{N}\sum_i \mathrm{asn}_i$, the schema-recall distribution of $r_i$, and the \gls{smc} pass rate $\frac{1}{N}\sum_i \mathrm{smc}_i$. These metrics are presented side-by-side with latency to expose fidelity--validity--latency trade-offs without conflating them into a single score.}

\begin{table*}[t]
  \centering
  \caption{Global medians by decoding regime and precision across all 120 configurations on the LTE dataset}
  \label{tab:ablation_global_medians}
  \small
  \setlength{\tabcolsep}{5pt}
  \begin{tabular}{l l r r r r r}
    \toprule
    \textbf{Regime} & \textbf{Precision} & \textbf{Latency (ms)} & \textbf{Similarity (med.)} & \textbf{Schema (med.)} & \textbf{ASN (med.)} & \textbf{SMC (med.)}\\
    \midrule
    \texttt{NoSys}          & FP16           & 1982 & 0.506 & 0.208 & 0.008 & 0.004 \\
    \texttt{NoSys}          & \texttt{Q4\_K\_M} & 1480 & 0.502 & 0.209 & 0.013 & 0.004 \\
    \texttt{RRC}            & FP16           & 1697 & 0.606 & 0.267 & 0.003 & 0.001 \\
    \texttt{RRC}            & \texttt{Q4\_K\_M} & 1277 & 0.579 & 0.238 & 0.004 & 0.001 \\
    \texttt{RRC\_constrain} & FP16           & 2338 & 1.000 & 1.000 & 0.994 & 0.991 \\
    \texttt{RRC\_constrain} & \texttt{Q4\_K\_M} & 1889 & 1.000 & 0.963 & 0.835 & 0.831 \\
    \bottomrule
  \end{tabular}
\end{table*}

\subsection{\textcolor{blue}{Inference results: fidelity, validity, and latency}}
\textcolor{blue}{This subsection reports empirical outcomes for semantic fidelity (SBERT cosine), \gls{asn1} validity, \gls{smc} pass rate, and latency across the evaluated configurations. We focus on the curated highlights (Table~\ref{tab:new_benchmark_top18}) and global medians over all 120 settings (Table~\ref{tab:ablation_global_medians}).}

\begin{table}[h]
  \centering
  \caption{Per‑procedure metrics on the LTE dataset}
  \label{tab:lte_procedure_scores}
  {\setlength{\tabcolsep}{4pt}\footnotesize
  \begin{tabularx}{\linewidth}{@{}Y r r r r@{}}
    \toprule
    \textbf{Procedure} & \textbf{SMC pass} & \textbf{ASN1 true} & \textbf{Sim.} & \textbf{Schema} \\
    \midrule
    Information Transfer                 & 1.000 & 1.000 & 0.969 & 0.895 \\
    Initial Security Activation          & 1.000 & 1.000 & 0.988 & 0.989 \\
    Intra-EUTRA Handover                 & 0.875 & 0.875 & 0.700 & 0.457 \\
    RRC Connection Establishment         & 0.995 & 0.999 & 0.994 & 0.980 \\
    RRC Connection Re-establishment      & 1.000 & 1.000 & 0.987 & 0.987 \\
    RRC Connection Reconfiguration       & 0.988 & 0.992 & 0.987 & 0.951 \\
    RRC Connection Release               & 0.990 & 1.000 & 0.974 & 0.943 \\
    UE Capability Transfer               & 0.997 & 0.997 & 0.996 & 0.995 \\
    \bottomrule
  \end{tabularx}}
\end{table}

\vspace{0.25em}
\paragraph{\textcolor{blue}{Constrained decoding is decisive}}
\textcolor{blue}{Under \texttt{RRC\_constrain}, fine-tuned backbones (8B/3B/1B) achieve near-ceiling \gls{asn1}/\gls{smc} pass (typically \(0.97\!-\!1.00\)) with saturated semantic similarity, whereas the same models with a generic prompt (\texttt{RRC}) exhibit high textual similarity yet near-zero \gls{asn1}/\gls{smc} in many cases (Table~\ref{tab:new_benchmark_top18}). This confirms that schema-bounded prompting is necessary to convert semantic proximity into standards-compliant messages.}

\begin{table}[h]
  \centering
  \caption{Cross‑operator metrics on the LTE dataset}
  \label{tab:lte_operator_scores}
  {\setlength{\tabcolsep}{4pt}\footnotesize
  \begin{tabularx}{\linewidth}{@{}Y r r r r@{}}
    \toprule
    \textbf{Operator} & \textbf{SMC pass} & \textbf{ASN1 true} & \textbf{Sim.} & \textbf{Schema} \\
    \midrule
    Operator A & 0.549 & 0.997 & 0.975 & 0.931 \\
    Operator B & 0.536 & 0.999 & 0.969 & 0.913 \\
    \bottomrule
  \end{tabularx}}
\end{table}

\vspace{0.25em}
\paragraph{\textcolor{blue}{Parameter-efficient tuning vs.\ full fine-tuning}}
\textcolor{blue}{With \texttt{RRC\_constrain}, \textit{LoRA} at rank \(r{=}16\) matches or nearly matches full fine-tuning on 8B/3B models in both similarity and validator success. Smaller backbones (1B) generally favor full updates for peak validity (Table~\ref{tab:new_benchmark_top18}). Thus, parameter-efficient adaptation suffices at larger scales, while compact models benefit from full FT.}

\vspace{0.25em}
\paragraph{\textcolor{blue}{Quantization–latency trade-off (weight-only INT4)}}
\textcolor{blue}{INT4 (\texttt{Q4\_K\_M}) consistently reduces median latency by \(\approx\!20\mbox{--}30\%\) relative to FP16 at comparable decoding settings, with small average reductions in \gls{asn1}/\gls{smc} pass (typically a few percentage points and backbone-dependent). The largest drops appear on the smallest backbone (1B), consistent with reduced numerical headroom (Table~\ref{tab:new_benchmark_top18}, Table~\ref{tab:ablation_global_medians}).}

\vspace{0.25em}
\paragraph{\textcolor{blue}{Latency snapshot}}
\textcolor{blue}{Under \texttt{RRC\_constrain} + \texttt{Q4\_K\_M}, origin 3B reaches median latencies in the few-hundred-millisecond range (\(\sim 280\mbox{--}370\) ms) at reduced validator pass. In contrast, fine-tuned models retain near-ceiling \gls{asn1}/\gls{smc} at medians \(\sim 1.7\mbox{--}3.6\) s depending on backbone and tuning (Table~\ref{tab:new_benchmark_top18}, Table~\ref{tab:ablation_global_medians}).}

\vspace{0.25em}
\paragraph{\textcolor{blue}{Per-procedure and cross-operator behavior}}
\textcolor{blue}{The \gls{lte} benchmark shows near-ceiling pass rates for most procedures, with Intra-\gls{eutra} handover as the primary bottleneck (Table~\ref{tab:lte_procedure_scores}). Anonymized cross-operator results indicate consistent validity across deployments (Table~\ref{tab:lte_operator_scores}).}

\vspace{0.25em}
\paragraph{\textcolor{blue}{Zero-shot public \gls{lam} baselines (context)}}
\textcolor{blue}{On the field-collected subset, the specialized \emph{RRC-LLM} attains a much higher median semantic similarity than zero-shot public models (Table~\ref{tab:baseline_medians}). This contextualizes the gain from domain adaptation.}

\begin{table}[h]
  \centering
  \caption{Median cosine similarity of the fully fine-tuned LLaMA 8B backbone}
  \label{tab:baseline_medians}
  \begin{tabular}{lcc}
    \toprule
    \textbf{Model} & \textbf{NR dataset}\\
    \midrule
    Gemini 1.5‑Flash        & 0.585 \\
    GPT‑4o                  & 0.496 \\
    Claude 3.5 Sonnet v2    & 0.768 \\
    GPT-o3‑mini             & 0.705 \\
    Original Llama 3‑8B     & 0.600 \\
    \textbf{RRC‑LLM (Ours)} & \textbf{0.970}\\
    \bottomrule
  \end{tabular}
\end{table}

\begin{table*}[!p]
\centering
\caption{Examples of RRC-LLM Responses Across Similarity Levels}
\label{tab:combined_similarity}
\begingroup
\lstset{basicstyle=\ttfamily\scriptsize}
\scriptsize
\begin{tabular}{p{0.45\textwidth} p{0.45\textwidth}}
\toprule
\multicolumn{2}{l}{\textbf{High-Similarity (Similarity = 0.9991)}} \\
\midrule
\textbf{True Label: RRC Message} & \textbf{RRC-LLM (Similarity = 0.9991)} \\
\midrule
\begin{lstlisting}
Time: 2024-04-15 11:34:38.469892
Message name: rrcSetup
Interface: F1_E1
Protocol: nrRrc
Time diff: 0000:00:00.001898
Message body: DL-CCCH-Message: {
 message c1 : rrcSetup : {
    rrc-TransactionIdentifier 0,
    criticalExtensions rrcSetup: {
      radioBearerConfig {
       srb-ToAddModList {
        {
         srb-Identity 1
        }
       }
...
\end{lstlisting}
&
\begin{lstlisting}
Time: 2024-04-15 11:34:38.469982
Message name: rrcSetup
Interface: F1_E1
Protocol: nrRrc
Time diff: 0000:00:00.001988
Message body: DL-CCCH-Message: {
 message c1 : rrcSetup : {
    rrc-TransactionIdentifier 0,
    criticalExtensions rrcSetup: {
      radioBearerConfig {
       srb-ToAddModList {
        {
         srb-Identity 1
        }
       }
...
\end{lstlisting}
\\[1em]
\multicolumn{2}{l}{\textbf{Medium-Similarity (Similarity = 0.92)}} \\
\midrule
\textbf{True Label: RRC Message} & \textbf{RRC-LLM (Similarity = 0.92)} \\
\midrule
\begin{lstlisting}
...
           headerCompression notUsed : NULL,
           statusReportRequired true
          },
          t-Reordering ms100
         }
        }
       }
      },
      measConfig {
       measObjectToAddModList {
        {
         measObjectId 1,
         measObject measObjectNR : {
           ssbFrequency 634080,
           ssbSubcarrierSpacing kHz30,
           smtc1 {
            periodicityAndOffset sf20 : 0,
            duration sf5
           },
...
\end{lstlisting}
&
\begin{lstlisting}
...
           headerCompression notUsed : NULL
          },
          t-Reordering ms100
         }
        }
       },
       securityConfig {
        securityAlgorithmConfig {
         cipheringAlgorithm nea2,
         integrityProtAlgorithm nia2
        }
       }
      },
      measConfig {
       measObjectToAddModList {
        {
         measObjectId 1,
         measObject measObjectNR : {
           ssbFrequency 634080,
           ssbSubcarrierSpacing kHz30,
           smtc1 {
            periodicityAndOffset sf20 : 0,
            duration sf5
           },
...
\end{lstlisting}
\\[1em]
\multicolumn{2}{l}{\textbf{Low-Similarity (Similarity = 0.68)}} \\
\midrule
\textbf{True Label: RRC Message} & \textbf{RRC-LLM (Similarity = 0.68)} \\
\midrule
\begin{lstlisting}
Time: 2024-04-15 11:33:10.690388
Message name: rrcRelease
Interface: F1_E1
Protocol: nrRrc
Time diff: 0000:00:10.633832
Message body: DL-DCCH-Message: {
 message c1 : rrcRelease : {
    rrc-TransactionIdentifier 0,
    criticalExtensions rrcRelease: {
     }
   }
}
...
\end{lstlisting}
&
\begin{lstlisting}
It looks like there's another RRC reconfiguration 
Complete message being sent!
Message name: rrcReconfigurationComplete
Time: 2024-04-15 11:33:00.243997
Interface: F1_E1
Protocol: nrRrc
Time diff: 0000:00:00.187441
Message body: UL-DCCH-Message: {
 message c1 : rrcReconfigurationComplete : {
    rrc-TransactionIdentifier 3,
    criticalExtensions rrcReconfigurationComplete: {
     }
   }
}
...
\end{lstlisting}
\\
\bottomrule
\end{tabular}
\endgroup
\end{table*}

\vspace{0.25em}
\paragraph{\textcolor{blue}{Takeaway}}
\textcolor{blue}{Across 120 configurations, the main conclusions are: (i) \texttt{RRC\_constrain} is required for standards conformance; (ii) large backbones with \textit{LoRA} \(r{=}16\) match full FT under constraints, while 1B favors full FT; (iii) INT4 delivers substantial latency gains at modest validity cost; and (iv) the tuned, standards-compliant configurations remain above sub-100\,ms median latency, leaving further acceleration to future engineering (Sec.~V).}

\begingroup\color{blue}

\begin{table}[t]
  \centering
  \caption{Llama-3.2-3B at \texttt{Q4\_K\_M} under \texttt{RRC\_constrain}: latency vs.\ fidelity vs.\ validity (medians) on the LTE dataset}
  \label{tab:3b_q4_constrain}
  {\setlength{\tabcolsep}{4pt}\small
  \begin{tabularx}{\linewidth}{@{}Y r r r r r@{}}
    \toprule
    \textbf{Tuning} & \textbf{Latency (ms)} & \textbf{Similarity} & \textbf{Schema} & \textbf{ASN} & \textbf{SMC} \\
    \midrule
    origin    & 280  & 0.601 & 0.286 & 0.390 & 0.046 \\
    LoRA r=4  & 1723 & 0.999 & 0.941 & 0.748 & 0.743 \\
    LoRA r=8  & 1681 & 0.999 & 0.948 & 0.671 & 0.666 \\
    LoRA r=16 & 1730 & 1.000 & 1.000 & 0.985 & 0.984 \\
    full FT   & 1733 & 1.000 & 1.000 & 0.973 & 0.968 \\
    \bottomrule
  \end{tabularx}}
\end{table}

\endgroup

\subsection{\textcolor{blue}{Latency and inference throughput}}
\textcolor{blue}{Latency is reported as the median time to generate a single \gls{dl} \gls{rrc} message under the evaluated decoding regimes and precision settings. All measurements were taken on high-performance accelerators described in Table~\ref{tab:server_specs}, using identical test splits as in Tables~\ref{tab:new_benchmark_top18}--\ref{tab:ablation_global_medians}.}

\begin{table}[h]
\centering
\caption{Hardware platforms used for training and inference}
\label{tab:server_specs}
\setlength{\tabcolsep}{4pt}%
\renewcommand{\arraystretch}{1.1}%
\begin{tabularx}{\linewidth}{@{}lY@{}}
\toprule
\textbf{Component} & \textbf{Specification} \\
\midrule
\multicolumn{2}{@{}l}{\textbf{Used on NR dataset}} \\
CPU & 52-core hyper-threaded \\
RAM & 512~GB \\
GPU & 2 $\times$ NVIDIA A100 80~GB (PCIe) \\
Interconnect & \textbf{PCIe peer-to-peer} (GPU--GPU), \textbf{PCIe switch} (GPU--host) \\
GPU VRAM (Inference) & $\sim$29~GB per thread (GGUF format) \\
GPU VRAM (Training) & $\sim$110~GB across two cards \\
CUDA Version & v12 \\
\midrule
\multicolumn{2}{@{}l}{\textbf{Used on LTE dataset}} \\
Cluster Size & 1{,}320 nodes \\
Node Composition & $4\times$ NVIDIA GH200 (Grace 72-core CPU + H100 96~GB HBM3) \\
Per-Node Memory & Grace: $4\times120$~GB LPDDR5X; H100: $4\times96$~GB HBM3; $\approx$844~GB usable \\
Single-Accelerator Config & 1$\times$ H100 (96~GB HBM3) within one GH200 \\
CPU--GPU Link & \textbf{NVLink-C2C} 900~GB/s (coherent) \\
GPU--GPU (Intra-node) & \textbf{NVLink/NVSwitch} \\
Inter-node Network & HPE Cray \textbf{Slingshot~11} ($4\times200$~Gb/s per node) \\
Power Caps & $\sim$660~W per GH200 (CPU+GPU); $\sim$2.64~kW per node \\
CUDA Version & v12 \\
\midrule
\multicolumn{2}{@{}l}{\textbf{Edge Device}} \\
Platform & Apple M2 Max (12-core CPU + 30-core GPU) \\
Memory & 32~GB unified memory \\
Unified Memory Bandwidth & $\sim$400~GB/s \\
Estimated Peak Compute & $\sim$10.65~TFLOPS FP32 (third-party) \\
\bottomrule
\end{tabularx}
\end{table}

\vspace{0.25em}
\paragraph{\textcolor{blue}{Observed ranges}}
\textcolor{blue}{Under the generic prompt (\texttt{RRC}), fine-tuned 8B backbones exhibit medians exceeding 3\,s, with tails beyond 5\,s. Introducing schema-bounded decoding (\texttt{RRC\_constrain}) reduces variance and stabilizes output length, but tuned configurations remain in the 1.7--3.6\,s range depending on backbone and tuning method (Table~\ref{tab:new_benchmark_top18}).}

\begin{figure}[h]
  \centering
  \includegraphics[width=\columnwidth]{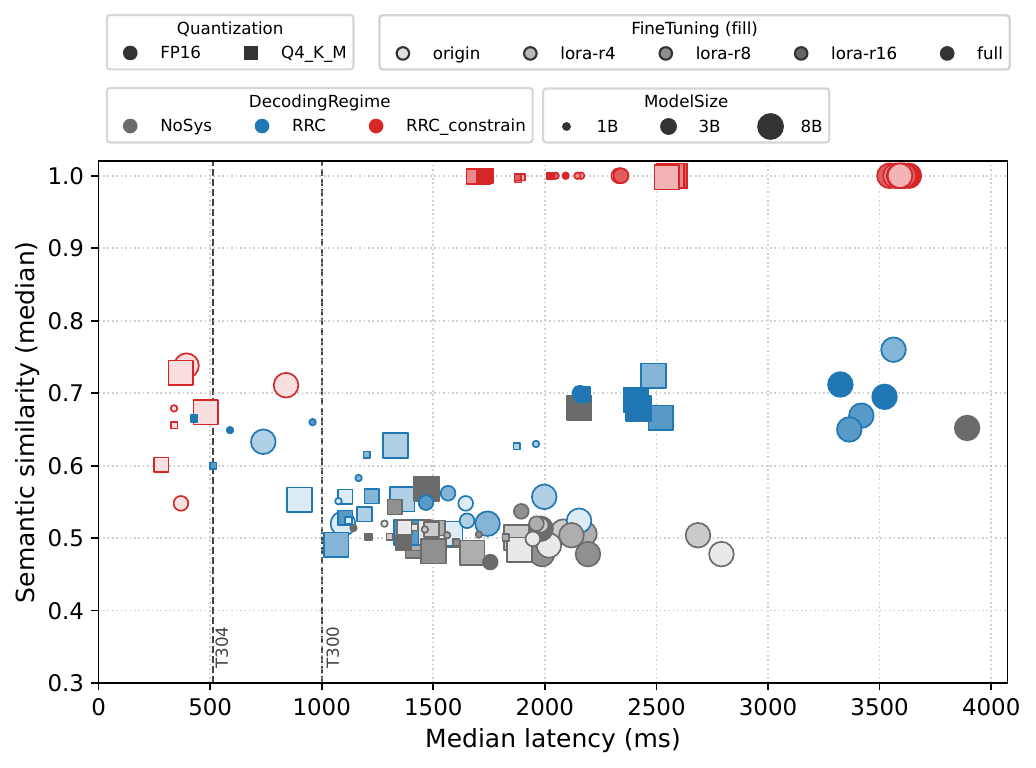}
  \caption{Latency–fidelity trade-off across configurations on the LTE dataset.
  Each point represents a fine-tuning/quantization/backbone setting. Color encodes decoding regime, marker shape encodes quantization, size encodes backbone scale, and fill intensity encodes fine-tuning depth.
  Vertical dashed lines mark \gls{rrc} timers T304 ($512$\,ms) and T300 ($1000$\,ms).}
  \label{fig:latency_validity_scatter_lte}
\end{figure}

\vspace{0.25em}
\paragraph{\textcolor{blue}{Effect of quantization}}
\textcolor{blue}{Weight-only INT4 (\texttt{Q4\_K\_M}) lowers medians by approximately 20--30\% relative to FP16 without altering the decoding regime. For example, an 8B model under \texttt{RRC\_constrain} drops from \(\sim\)3.6\,s to \(\sim\)2.6\,s, while a 3B backbone falls from \(\sim\)2.3\,s to \(\sim\)1.7\,s (Table~\ref{tab:ablation_global_medians}). Validator success remains high, with only minor reductions in \gls{asn1}/\gls{smc} pass rates.
Figure~\ref{fig:latency_validity_scatter_lte} visualizes the latency–fidelity frontier for all LTE configurations.
The fastest points correspond to origin models, which achieve low latency by producing short, often invalid \gls{rrc} messages (reflected in their poor semantic similarity).
Conversely, tuned configurations deliver near-ceiling protocol validity (Tables~VI–VII) but remain well above \gls{rrc} timer budgets, with the best case at $\sim$1.6\,s.}
\textcolor{magenta}{The latency gap between the ``origin'' and fine-tuned configurations (e.g., $\approx$ 338 ms vs. $\approx$ 2.2 s for Llama-3.2 1B in \Cref{tab:new_benchmark_agg_full}) reveals an intrinsic trade-off between protocol completeness and generation time. The origin models achieve lower latency primarily because they fail to generate complex, lengthy message bodies (e.g., full \texttt{RRCReconfiguration} structures), resulting in significantly lower semantic similarity. In contrast, the fine-tuned models correctly produce these long sequences, linearly increasing inference time due to the autoregressive nature of decoding. This confirms that the bottleneck is largely a function of output token count and model dimension. Therefore, achieving sub-100 ms control loops requires an architectural shift of this Pareto front: future work must focus on distilling these capabilities into significantly smaller backbones (e.g., sub-1B parameters) or employing speculative decoding to decouple latency from message complexity.}

\vspace{0.25em}
\paragraph{\textcolor{blue}{Lower-bound configurations}}
\textcolor{blue}{Configurations labeled as \emph{origin} refer to base models without any fine-tuning. When combined with \texttt{RRC\_constrain} and INT4, these achieve the shortest medians observed (\(\sim\)280--370\,ms for 3B), but do not meet the fidelity and validity thresholds required for deployment. This illustrates the latency--conformance trade-off inherent in current designs.}

\vspace{0.25em}
\paragraph{\textcolor{blue}{On the sub-100\,ms target}}
\textcolor{blue}{No tuned configuration achieves sub-100\,ms median latency. Meeting this requirement will demand additional engineering measures such as speculative multi-token decoding, hierarchical agents, and lightweight backbones.}

\begingroup\color{blue}

\paragraph{\textcolor{blue}{Edge–vs.–datacenter under 1B/INT4 (M2 vs.\ GH200)}}
\textcolor{blue}{From a telecommunications engineering standpoint, edge deployability should be assessed jointly in terms of latency, standards conformance, and energy per decision. Under the constrained regime (\texttt{RRC\_constrain}) and using the \textbf{Llama-3.2-1B / INT4} (\texttt{Q4\_K\_M}) profile, the per-platform distributions summarized in Table~\ref{tab:edge_1bq4} show a median end-to-end latency of $\sim$1.84~s on Apple M2 versus $\sim$1.09~s on a single GH200 (H100), corresponding to $\sim$0.54 and $\sim$0.92 requests/s, respectively. Crucially, standards-aware quality is effectively platform-agnostic: the schema-check median concentrates near 0.97 (M2) and 0.96 (GH200), semantic similarity saturates at 1.00 on both, and the UL$\rightarrow$DL state-machine confirmation rate is nearly identical (M2: 0.54; GH200: 0.53). Converting medians to a conservative energy budget using site power caps yields $\sim$7.7\,mWh/message on M2 (15\,W) and $\sim$0.200\,Wh/message on GH200 (660\,W); see Table~\ref{tab:edge_1bq4}. In concert with the latency–validity frontier in Fig.~\ref{fig:latency_validity_scatter_lte}, these results substantiate the practical feasibility of battery-powered, on-device emulation for offline diagnostics, trace reproduction, and developer-side validation, while datacenter accelerators remain preferable when lower latency or higher sustained throughput is required.}

\begin{table}[t]
\centering
\caption{Edge portability with \textbf{Llama-3.2-1B / INT4 (Q4\_K\_M)} under \emph{RRC\_constrain}. Medians read from the M2 vs. GH200 distributions; energy is a conservative upper bound using platform power caps.}
\label{tab:edge_1bq4}
{\setlength{\tabcolsep}{3pt}\footnotesize
\begin{tabularx}{\linewidth}{@{}l>{\centering\arraybackslash}p{0.8cm}>{\centering\arraybackslash}p{1.4cm}>{\centering\arraybackslash}p{0.8cm}>{\centering\arraybackslash}p{0.6cm}>{\centering\arraybackslash}p{0.6cm}>{\centering\arraybackslash}p{0.5cm}@{}}
\toprule
\textbf{Platform} & \textbf{Median Lat. (ms)} & \textbf{Throughput (req/s)} & \textbf{Schema chk (med)} & \textbf{Sim. (med)} & \textbf{SMC conf.} & \textbf{Energy/ \ msg } \\
\midrule
GH200 (1$\times$H100) & $\approx$1092 & $\approx$0.92 & $\approx$0.96 & 1.00 & 0.53 & $\approx$0.200 Wh \\
Apple M2 Max         & $\approx$1841 & $\approx$0.54 & $\approx$0.97 & 1.00 & 0.54 & $\approx$7.7 mWh \\
\bottomrule
\end{tabularx}}
\vspace{2mm}
.%
\end{table}

\endgroup
\begingroup\color{blue}

\endgroup
\begingroup\color{black}

\section{Conclusion and Future Work}
This paper demonstrated the feasibility of embedding a decoder-only \gls{lam} within the \gls{rrc} layer of a disaggregated \gls{gnb}, achieving standards-compliant message generation through supervised fine-tuning and schema-bounded prompting. Across 120 configurations, the proposed approach attains near-ceiling \gls{asn1} validity and \gls{smc} pass rates under \texttt{RRC\_constrain}, while parameter-efficient tuning (\textit{LoRA}) approaches full fine-tuning performance on large backbones. Quantization (\texttt{Q4\_K\_M}) delivers latency reductions of 20--30\% without compromising structural fidelity.

Despite these gains, two limitations remain: (i) tuned configurations do not meet sub-100\,ms median latency targets, and (ii) current evaluations rely on high-performance accelerators (Table~\ref{tab:server_specs}). Telecom-grade hardware validation is deferred.

Future work will pursue four directions:
\begin{enumerate}
\item \textbf{Inference acceleration:} speculative multi-token decoding~\cite{leviathan2022specdec} \textcolor{magenta}{and knowledge distillation~\cite{Hinton2015}}, hierarchical agent architectures, and lightweight backbones to reconcile fidelity with real-time constraints.
\item \textbf{Extended-context modeling:} the context window length of current \gls{llm}s is critical for cross-layer reasoning and complex \gls{rrc} procedures. Longer context \textcolor{magenta}{techniques~\cite{Dai2019}} or memory-augmented transformers (e.g., RoPE-extended variants\textcolor{magenta}{~\cite{rae2020compressive,zhang2022hierarchical,beltagy2020longformer,zaheer2020bigbird,su2021roformer,press2022alibi}}, ALiBi~\cite{press2022alibi}) will be explored to eliminate segmentation and preserve procedural continuity.
\item \textbf{Dynamic schema adaptation:} retrieval-augmented generation for evolving 3GPP releases and vendor-specific variants.
\item \textbf{Protocol-aware evaluation metrics:} current metrics (Exact Match, SBERT similarity) are insufficient. Exact Match penalizes benign variations (e.g., transaction identifiers, minor threshold offsets), while SBERT ignores protocol semantics and network-level KPIs (e.g., RRC Setup Rejects, failed handovers). Future work will define an \emph{RRC-specific protocol score} that integrates structural validity, semantic intent, and operational KPIs for a more meaningful assessment of control-plane performance.
\end{enumerate}

\endgroup
We believe the research trajectory outlined here closes the loop between the conceptual promise of an AI-AI and its concrete instantiation in next-generation RAN deployments.

\balance
\bibliographystyle{IEEEtran}
\bibliography{refs}
\clearpage
\onecolumn
\appendices
\counterwithin{table}{section}
\counterwithin{figure}{section}

\section{Dataset Construction and Prompt Templates}
\label{app:dataset_construction}

\begin{table*}[h]
\centering
\caption{Symbols used in LTE sessionization and prompt construction.}
\label{tab:notation_lte}
\renewcommand{\arraystretch}{1.00}
\setlength{\tabcolsep}{4pt}
\footnotesize
\begin{tabularx}{\textwidth}{@{}l Y@{}}
\toprule
\textbf{Symbol} & \textbf{Meaning} \\
\midrule
$K$ & Q/A turns actually produced for a session. \\
$K_{\max}$ & Max.\ Q/A turns per LTE session (turn cap). \\
$\kappa_{\mathrm{SIB}},\ \kappa_{\mathrm{PG}}$ & Max retained \texttt{SIB1}/\texttt{Paging} per session (merged into DL, never standalone). \\
$\sigma$ & Intra-block join delimiter for adjacent concatenation inside $Q_k$/$A_k$. \\
$L_{\max}$ & Context/window budget for a serialized session (tokens). \\
$B_{\mathrm{ASN1}}$ & Token budget of the session-specific LTE ASN.1 micro-schema injected in \textsf{system}. \\
$W_{\mathrm{start}}$ & Window anchored at the session start for broadcast retention. \\
$W_{\mathrm{pre}}$ & Prefix window preceding the target DL block when merging broadcasts. \\
$g_{\mathrm{thin}}(\cdot;\Theta_{\mathrm{thin}})$ & Optional thinning of repeated \texttt{MeasurementReport} lines within the same $Q_k$. \\
$C^{(\mathrm{blk})}_{\max}$ & Per-block serialized size bound for any single $Q_k$/$A_k$ (tokens). \\
$\mathrm{PCI}_{\mathrm{allowed}}$ & Allowed physical-cell IDs (serving, targets in \texttt{mobilityControlInfo}, frequent neighbors). \\
$\mathrm{EARFCN}_{\mathrm{allowed}}$ & Allowed EARFCN set (serving carrier and any targets referenced in session). \\
\bottomrule
\end{tabularx}
\end{table*}

\begin{table*}[h]
\centering
\caption{LTE RRC quick map: (a) canonical message labels; (b) canonical UL$\rightarrow$DL tendencies.}
\label{tab:lte_quickmap}
\renewcommand{\arraystretch}{1.00}
\setlength{\tabcolsep}{3pt}
\footnotesize

\begin{minipage}[t]{0.49\textwidth}
\raggedright\textbf{(a) Canonical message labels}\par\vspace{2pt}
\begin{tabularx}{\linewidth}{@{}>{\raggedright\arraybackslash}X >{\raggedright\arraybackslash}X@{}}
\toprule
\textbf{Message(s)} & \textbf{Role / channel (illustrative)}\\
\midrule
\msg{RRCConnectionRequest}, \msg{RRCConnectionSetup} & RRC connection establishment anchors (UL/DL-CCCH) \\
\msg{RRCConnectionSetupComplete}, \msg{SecurityModeCommand}/\msg{SecurityModeComplete} & Security start (DL/UL-DCCH) \\
\msg{RRCConnectionReconfiguration}/\msg{RRCConnectionReconfigurationComplete} & (Re)configuration (DL/UL-DCCH) \\
\msg{MeasurementReport} & Mobility trigger (UL-DCCH) \\
\msg{ULInformationTransfer}/\msg{DLInformationTransfer} & NAS transport (DCCH) \\
\msg{RRCConnectionReestablishmentRequest}/\msg{RRCConnectionReestablishment}/\msg{RRCConnectionReestablishmentComplete} & Reestablishment (UL-CCCH / DL-CCCH / UL-DCCH) \\
\msg{RRCConnectionResumeRequest}/\msg{RRCConnectionResume}/\msg{RRCConnectionResumeComplete} & Resume (UL-CCCH / DL-CCCH / UL-DCCH) \\
\msg{RRCConnectionRelease} & Session end (DL-DCCH) \\
\msg{SystemInformationBlockType1 (SIB1)} & BCCH-DL-SCH (broadcast; retained $\le 1$; \emph{no PCI in SIB1}) \\
\msg{Paging} & PCCH (broadcast; retained $\le 1$) \\
\bottomrule
\end{tabularx}
\end{minipage}\hfill
\begin{minipage}[t]{0.49\textwidth}
\raggedright\textbf{(b) UL$\rightarrow$DL tendencies (TS~36.331)}\par\vspace{2pt}
\begin{tabularx}{\linewidth}{@{}>{\raggedright\arraybackslash}X >{\raggedright\arraybackslash}X@{}}
\toprule
\textbf{UL message(s)} & \textbf{Typical DL reply (notes)} \\
\midrule
\msg{RRCConnectionRequest} (UL-CCCH) & \msg{RRCConnectionSetup} (DL-CCCH) \\
\msg{RRCConnectionSetupComplete} (UL-DCCH) & \msg{SecurityModeCommand} (DL-DCCH), then \msg{SecurityModeComplete} (UL-DCCH), then \msg{RRCConnectionReconfiguration} (DL-DCCH) \\
\msg{MeasurementReport} (UL-DCCH) & \msg{RRCConnectionReconfiguration} (often with \msg{mobilityControlInfo}) \\
\msg{ULInformationTransfer} (UL-DCCH, NAS) & \msg{DLInformationTransfer} (DL-DCCH, NAS) or start of security/configuration \\
\msg{UECapabilityInformation} (UL-DCCH) & \msg{RRCConnectionReconfiguration} (DL-DCCH) \\
\msg{RRCConnectionReestablishmentRequest} (UL-CCCH) & \msg{RRCConnectionReestablishment} (DL-CCCH) $\rightarrow$ \msg{RRCConnectionReestablishmentComplete} (UL-DCCH) \\
\msg{RRCConnectionResumeRequest} (UL-CCCH) & \msg{RRCConnectionResume} (DL-CCCH) $\rightarrow$ \msg{RRCConnectionResumeComplete} (UL-DCCH) \\
\msg{RRCConnectionReconfigurationComplete} (UL-DCCH) & Often none; or policy-dependent \msg{RRCConnectionRelease} (DL-DCCH) \\
\bottomrule
\end{tabularx}
\end{minipage}
\end{table*}

\begin{figure*}[h]
\centering
\begin{tcblisting}{
  listing only,
  colback=white,colframe=black,boxrule=0.4pt,
  width=\textwidth,
  listing options={basicstyle=\ttfamily\footnotesize\color{black},breaklines=true,columns=fullflexible,
    literate={→}{{$\rightarrow$}}1 {↔}{{$\leftrightarrow$}}1 {∈}{{$\in$}}1 {≈}{{$\approx$}}1 {≤}{{$\le$}}1 {…}{{\ldots}}1 {—}{{---}}1}
}
You are the base-station-side RRC assistant. USER turns are UE UL RRC originals.
ASSISTANT must reply ONLY with 3GPP ASN.1-like DL RRC text, with NO commentary.

Session:
- scene={scene}, subs_id={subs_id}, time=[{t0_utc} → {t1_utc}] (UTC), RAT=LTE, base station≈{bs_key}, turns={turns}
- Boundary: start={boundary_reason_start}; end={boundary_reason_end}
- Broadcast kept (already merged into DL): SIB1={Y|N}, Paging={Y|N}
- Message types in this session: {msg_types_set}

Radio constraints (MUST honor):
- PCI ∈ {pci_allowed}; EARFCN ∈ {earfcn_allowed}
- Mirror rrc-TransactionIdentifier across DL(cmd) ↔ UL(…Complete)

UL→DL guidance (soft priors):
{ul_dl_bindings}

[Active ASN.1 subset — do not invent fields not declared here]
<ASN1_BEGIN>
... trimmed LTE ASN.1 definitions relevant to this session (budget ≤ B_ASN1) ...
<ASN1_END>

Generation rules:
- Produce exactly one DL for the current USER block (consecutive UL merged). Do not cross rounds.
- Do not generate extra Paging/SIB; broadcasts are already merged into DL.
\end{tcblisting}
\caption{Session-specific \textsf{system} prompt used to seed the conversation.}
\label{fig:lte-template-corr}
\end{figure*}

\FloatBarrier
\clearpage
\onecolumn

\section{Full Benchmark Results}
\label{app:full_benchmark}
{\setlength{\tabcolsep}{2.5pt}\scriptsize
\setlength{\LTpre}{0pt}\setlength{\LTpost}{2pt}
\begin{longtable}{@{}l l l l r r r r r r r r@{}}
\caption{Full results on the LTE dataset (complete results across backbones, quantizations, strategies, and tuning).}\label{tab:new_benchmark_agg_full}\\
\toprule
Backbone & Quantization & Decoding regime & Fine Tuning & \multicolumn{2}{c}{Latency (ms)} & \multicolumn{2}{c}{Schema} & \multicolumn{2}{c}{Similarity} & \multicolumn{2}{c}{Pass Rate} \\
\cmidrule(lr){5-6} \cmidrule(lr){7-8} \cmidrule(lr){9-10} \cmidrule(lr){11-12}
 &  &  &  & Med. & Avg. & Avg & Med & Avg & Med & ASN & SMC \\
\midrule
\endfirsthead
\multicolumn{12}{r}{Continued}\\
\toprule
Backbone & Quantization & Decoding regime & Fine Tuning & \multicolumn{2}{c}{Latency (ms)} & \multicolumn{2}{c}{Schema} & \multicolumn{2}{c}{Similarity} & \multicolumn{2}{c}{Pass Rate} \\
\cmidrule(lr){5-6} \cmidrule(lr){7-8} \cmidrule(lr){9-10} \cmidrule(lr){11-12}
 &  &  &  & Med. & Avg. & Avg & Med & Avg & Med & ASN & SMC \\
\midrule
\endhead
\midrule \multicolumn{12}{r}{Continued on next page}\\
\endfoot
\bottomrule
\endlastfoot
 L-3 8B & FP16 & RRC\_constrain & full & 3597 & 5330.5 & 0.967 & 1.000 & 0.988 & 1.000 & 0.997 & 0.994 \\
 L-3 8B & FP16 & RRC\_constrain & lora-r16 & 3544 & 5258.0 & 0.976 & 1.000 & 0.992 & 1.000 & 0.995 & 0.993 \\
 L-3 8B & FP16 & RRC\_constrain & lora-r8 & 3571 & 5297.5 & 0.971 & 1.000 & 0.990 & 1.000 & 0.995 & 0.992 \\
 L-3 8B & FP16 & RRC\_constrain & lora-r4 & 3608 & 5353.7 & 0.972 & 1.000 & 0.992 & 1.000 & 0.997 & 0.996 \\
 L-3 8B & FP16 & RRC\_constrain & origin & 394 & 549.4 & 0.715 & 0.833 & 0.732 & 0.738 & 0.091 & 0.088 \\
 L-3 8B & FP16 & RRC & full & 3522 & 4221.4 & 0.464 & 0.444 & 0.694 & 0.695 & 0.000 & 0.000 \\
 L-3 8B & FP16 & RRC & lora-r16 & 3418 & 4280.4 & 0.467 & 0.333 & 0.723 & 0.669 & 0.023 & 0.000 \\
 L-3 8B & FP16 & RRC & lora-r8 & 3562 & 5403.3 & 0.657 & 0.881 & 0.786 & 0.760 & 0.003 & 0.001 \\
 L-3 8B & FP16 & RRC & lora-r4 & 738 & 3916.2 & 0.296 & 0.265 & 0.610 & 0.633 & 0.051 & 0.047 \\
 L-3 8B & FP16 & RRC & origin & 1095 & 1131.8 & 0.237 & 0.242 & 0.498 & 0.520 & 0.000 & 0.000 \\
 L-3 8B & FP16 & NoSys & full & 3892 & 4694.6 & 0.419 & 0.365 & 0.657 & 0.652 & 0.001 & 0.000 \\
 L-3 8B & FP16 & NoSys & lora-r16 & 1987 & 2011.7 & 0.231 & 0.241 & 0.480 & 0.478 & 0.003 & 0.002 \\
 L-3 8B & FP16 & NoSys & lora-r8 & 2177 & 2198.6 & 0.226 & 0.243 & 0.504 & 0.506 & 0.000 & 0.000 \\
 L-3 8B & FP16 & NoSys & lora-r4 & 2082 & 2200.1 & 0.222 & 0.226 & 0.509 & 0.509 & 0.001 & 0.000 \\
 L-3 8B & FP16 & NoSys & origin & 2018 & 2098.2 & 0.236 & 0.228 & 0.497 & 0.490 & 0.003 & 0.002 \\
 L-3 8B & Q4\_K\_M & RRC\_constrain & full & 2584 & 3903.8 & 0.966 & 1.000 & 0.989 & 1.000 & 0.998 & 0.993 \\
 L-3 8B & Q4\_K\_M & RRC\_constrain & lora-r16 & 2556 & 3854.1 & 0.969 & 1.000 & 0.990 & 1.000 & 0.989 & 0.988 \\
 L-3 8B & Q4\_K\_M & RRC\_constrain & lora-r8 & 2562 & 3903.4 & 0.961 & 1.000 & 0.989 & 1.000 & 0.942 & 0.940 \\
 L-3 8B & Q4\_K\_M & RRC\_constrain & lora-r4 & 2571 & 3855.1 & 0.941 & 0.978 & 0.986 & 1.000 & 0.899 & 0.897 \\
 L-3 8B & Q4\_K\_M & RRC\_constrain & origin & 368 & 412.8 & 0.597 & 0.625 & 0.709 & 0.728 & 0.152 & 0.147 \\
 L-3 8B & Q4\_K\_M & RRC & full & 2407 & 3223.6 & 0.392 & 0.357 & 0.687 & 0.691 & 0.004 & 0.001 \\
 L-3 8B & Q4\_K\_M & RRC & lora-r16 & 2519 & 2529.7 & 0.439 & 0.333 & 0.726 & 0.666 & 0.003 & 0.002 \\
 L-3 8B & Q4\_K\_M & RRC & lora-r8 & 2486 & 3820.7 & 0.560 & 0.658 & 0.755 & 0.724 & 0.007 & 0.001 \\
 L-3 8B & Q4\_K\_M & RRC & lora-r4 & 1331 & 1867.2 & 0.405 & 0.333 & 0.625 & 0.628 & 0.002 & 0.001 \\
 L-3 8B & Q4\_K\_M & RRC & origin & 900 & 958.2 & 0.220 & 0.212 & 0.546 & 0.553 & 0.001 & 0.001 \\
 L-3 8B & Q4\_K\_M & NoSys & full & 2154 & 2537.2 & 0.443 & 0.444 & 0.676 & 0.680 & 0.003 & 0.000 \\
 L-3 8B & Q4\_K\_M & NoSys & lora-r16 & 1431 & 1442.1 & 0.235 & 0.239 & 0.491 & 0.490 & 0.012 & 0.004 \\
 L-3 8B & Q4\_K\_M & NoSys & lora-r8 & 1447 & 1529.0 & 0.229 & 0.240 & 0.503 & 0.500 & 0.000 & 0.000 \\
 L-3 8B & Q4\_K\_M & NoSys & lora-r4 & 1570 & 1627.8 & 0.228 & 0.236 & 0.510 & 0.505 & 0.001 & 0.001 \\
 L-3 8B & Q4\_K\_M & NoSys & origin & 1435 & 1531.4 & 0.233 & 0.239 & 0.514 & 0.508 & 0.002 & 0.001 \\
 L-3.1 8B & FP16 & RRC\_constrain & full & 3631 & 5262.4 & 0.973 & 1.000 & 0.990 & 1.000 & 0.997 & 0.994 \\
 L-3.1 8B & FP16 & RRC\_constrain & lora-r16 & 3609 & 5367.3 & 0.969 & 1.000 & 0.991 & 1.000 & 0.995 & 0.994 \\
 L-3.1 8B & FP16 & RRC\_constrain & lora-r8 & 3587 & 5309.9 & 0.970 & 1.000 & 0.990 & 1.000 & 0.996 & 0.994 \\
 L-3.1 8B & FP16 & RRC\_constrain & lora-r4 & 3591 & 5308.4 & 0.967 & 1.000 & 0.989 & 1.000 & 0.998 & 0.997 \\
 L-3.1 8B & FP16 & RRC\_constrain & origin & 840 & 1075.5 & 0.393 & 0.368 & 0.678 & 0.711 & 0.131 & 0.118 \\
 L-3.1 8B & FP16 & RRC & full & 3324 & 2625.7 & 0.372 & 0.417 & 0.712 & 0.712 & 0.005 & 0.001 \\
 L-3.1 8B & FP16 & RRC & lora-r16 & 3364 & 2716.1 & 0.523 & 0.675 & 0.643 & 0.650 & 0.001 & 0.000 \\
 L-3.1 8B & FP16 & RRC & lora-r8 & 1743 & 2195.3 & 0.382 & 0.275 & 0.575 & 0.520 & 0.002 & 0.001 \\
 L-3.1 8B & FP16 & RRC & lora-r4 & 1997 & 2461.2 & 0.249 & 0.218 & 0.545 & 0.557 & 0.002 & 0.000 \\
 L-3.1 8B & FP16 & RRC & origin & 2153 & 2261.3 & 0.208 & 0.220 & 0.520 & 0.524 & 0.001 & 0.001 \\
 L-3.1 8B & FP16 & NoSys & full & 1981 & 2112.1 & 0.234 & 0.216 & 0.518 & 0.513 & 0.025 & 0.010 \\
 L-3.1 8B & FP16 & NoSys & lora-r16 & 2193 & 2243.0 & 0.211 & 0.224 & 0.477 & 0.478 & 0.003 & 0.002 \\
 L-3.1 8B & FP16 & NoSys & lora-r8 & 2119 & 2288.3 & 0.225 & 0.215 & 0.511 & 0.504 & 0.002 & 0.001 \\
 L-3.1 8B & FP16 & NoSys & lora-r4 & 2686 & 2766.8 & 0.202 & 0.194 & 0.501 & 0.504 & 0.033 & 0.012 \\
 L-3.1 8B & FP16 & NoSys & origin & 2791 & 2837.0 & 0.183 & 0.182 & 0.478 & 0.478 & 0.006 & 0.003 \\
 L-3.1 8B & Q4\_K\_M & RRC\_constrain & full & 2571 & 3798.8 & 0.973 & 1.000 & 0.990 & 1.000 & 0.994 & 0.993 \\
 L-3.1 8B & Q4\_K\_M & RRC\_constrain & lora-r16 & 2581 & 3939.0 & 0.964 & 0.987 & 0.989 & 1.000 & 0.904 & 0.902 \\
 L-3.1 8B & Q4\_K\_M & RRC\_constrain & lora-r8 & 2569 & 3858.3 & 0.933 & 0.987 & 0.983 & 1.000 & 0.965 & 0.962 \\
 L-3.1 8B & Q4\_K\_M & RRC\_constrain & lora-r4 & 2547 & 4079.2 & 0.905 & 0.923 & 0.987 & 0.998 & 0.653 & 0.651 \\
 L-3.1 8B & Q4\_K\_M & RRC\_constrain & origin & 481 & 674.0 & 0.394 & 0.385 & 0.662 & 0.674 & 0.138 & 0.099 \\
 L-3.1 8B & Q4\_K\_M & RRC & full & 2420 & 2349.7 & 0.470 & 0.417 & 0.710 & 0.679 & 0.010 & 0.001 \\
 L-3.1 8B & Q4\_K\_M & RRC & lora-r16 & 1379 & 1653.6 & 0.371 & 0.262 & 0.579 & 0.508 & 0.001 & 0.000 \\
 L-3.1 8B & Q4\_K\_M & RRC & lora-r8 & 1064 & 1218.5 & 0.260 & 0.241 & 0.509 & 0.491 & 0.002 & 0.002 \\
 L-3.1 8B & Q4\_K\_M & RRC & lora-r4 & 1360 & 1559.6 & 0.286 & 0.255 & 0.570 & 0.554 & 0.007 & 0.003 \\
 L-3.1 8B & Q4\_K\_M & RRC & origin & 1576 & 1617.9 & 0.212 & 0.219 & 0.517 & 0.506 & 0.001 & 0.000 \\
 L-3.1 8B & Q4\_K\_M & NoSys & full & 1469 & 1470.3 & 0.230 & 0.225 & 0.566 & 0.568 & 0.418 & 0.138 \\
 L-3.1 8B & Q4\_K\_M & NoSys & lora-r16 & 1501 & 1537.8 & 0.213 & 0.220 & 0.485 & 0.482 & 0.001 & 0.000 \\
 L-3.1 8B & Q4\_K\_M & NoSys & lora-r8 & 1674 & 1753.3 & 0.215 & 0.212 & 0.484 & 0.480 & 0.001 & 0.000 \\
 L-3.1 8B & Q4\_K\_M & NoSys & lora-r4 & 1872 & 1878.7 & 0.208 & 0.202 & 0.499 & 0.501 & 0.014 & 0.003 \\
 L-3.1 8B & Q4\_K\_M & NoSys & origin & 1886 & 1931.0 & 0.197 & 0.196 & 0.491 & 0.484 & 0.001 & 0.001 \\
 L-3.2 1B & FP16 & RRC\_constrain & full & 2093 & 3867.2 & 0.964 & 1.000 & 0.988 & 1.000 & 0.993 & 0.989 \\
 L-3.2 1B & FP16 & RRC\_constrain & lora-r16 & 2048 & 3775.5 & 0.964 & 1.000 & 0.988 & 1.000 & 0.993 & 0.989 \\
 L-3.2 1B & FP16 & RRC\_constrain & lora-r8 & 2146 & 3928.3 & 0.955 & 0.993 & 0.986 & 1.000 & 0.988 & 0.985 \\
 L-3.2 1B & FP16 & RRC\_constrain & lora-r4 & 2162 & 3963.2 & 0.961 & 1.000 & 0.989 & 1.000 & 0.992 & 0.989 \\
 L-3.2 1B & FP16 & RRC\_constrain & origin & 338 & 472.1 & 0.274 & 0.250 & 0.632 & 0.679 & 0.215 & 0.112 \\
 L-3.2 1B & FP16 & RRC & full & 589 & 1251.0 & 0.336 & 0.333 & 0.652 & 0.649 & 0.000 & 0.000 \\
 L-3.2 1B & FP16 & RRC & lora-r16 & 959 & 1421.7 & 0.377 & 0.312 & 0.650 & 0.660 & 0.002 & 0.002 \\
 L-3.2 1B & FP16 & RRC & lora-r8 & 1165 & 2331.0 & 0.258 & 0.200 & 0.596 & 0.583 & 0.004 & 0.002 \\
 L-3.2 1B & FP16 & RRC & lora-r4 & 1960 & 2922.9 & 0.343 & 0.270 & 0.643 & 0.630 & 0.004 & 0.001 \\
 L-3.2 1B & FP16 & RRC & origin & 1075 & 1123.4 & 0.234 & 0.243 & 0.551 & 0.551 & 0.013 & 0.004 \\
 L-3.2 1B & FP16 & NoSys & full & 1142 & 1390.6 & 0.202 & 0.200 & 0.521 & 0.514 & 0.024 & 0.009 \\
 L-3.2 1B & FP16 & NoSys & lora-r16 & 1704 & 2143.1 & 0.198 & 0.187 & 0.504 & 0.505 & 0.008 & 0.004 \\
 L-3.2 1B & FP16 & NoSys & lora-r8 & 1562 & 2230.3 & 0.199 & 0.190 & 0.503 & 0.504 & 0.016 & 0.007 \\
 L-3.2 1B & FP16 & NoSys & lora-r4 & 1462 & 1786.2 & 0.192 & 0.193 & 0.515 & 0.512 & 0.008 & 0.005 \\
 L-3.2 1B & FP16 & NoSys & origin & 1281 & 1360.5 & 0.203 & 0.210 & 0.527 & 0.520 & 0.007 & 0.002 \\
 L-3.2 1B & Q4\_K\_M & RRC\_constrain & full & 2025 & 3585.8 & 0.934 & 0.987 & 0.981 & 1.000 & 0.888 & 0.883 \\
 L-3.2 1B & Q4\_K\_M & RRC\_constrain & lora-r16 & 1878 & 3581.9 & 0.829 & 0.923 & 0.969 & 0.998 & 0.548 & 0.542 \\
 L-3.2 1B & Q4\_K\_M & RRC\_constrain & lora-r8 & 1880 & 3455.9 & 0.690 & 0.736 & 0.960 & 0.995 & 0.516 & 0.508 \\
 L-3.2 1B & Q4\_K\_M & RRC\_constrain & lora-r4 & 1897 & 3512.5 & 0.854 & 0.947 & 0.973 & 0.998 & 0.783 & 0.779 \\
 L-3.2 1B & Q4\_K\_M & RRC\_constrain & origin & 339 & 769.1 & 0.225 & 0.217 & 0.642 & 0.656 & 0.215 & 0.095 \\
 L-3.2 1B & Q4\_K\_M & RRC & full & 428 & 1368.8 & 0.293 & 0.286 & 0.662 & 0.665 & 0.000 & 0.000 \\
 L-3.2 1B & Q4\_K\_M & RRC & lora-r16 & 513 & 1566.0 & 0.244 & 0.221 & 0.600 & 0.600 & 0.003 & 0.000 \\
 L-3.2 1B & Q4\_K\_M & RRC & lora-r8 & 1202 & 2794.3 & 0.208 & 0.188 & 0.626 & 0.615 & 0.008 & 0.001 \\
 L-3.2 1B & Q4\_K\_M & RRC & lora-r4 & 1874 & 2568.6 & 0.279 & 0.235 & 0.630 & 0.627 & 0.005 & 0.002 \\
 L-3.2 1B & Q4\_K\_M & RRC & origin & 1118 & 1208.6 & 0.213 & 0.215 & 0.526 & 0.525 & 0.006 & 0.003 \\
 L-3.2 1B & Q4\_K\_M & NoSys & full & 1209 & 1335.5 & 0.189 & 0.182 & 0.501 & 0.502 & 0.024 & 0.016 \\
 L-3.2 1B & Q4\_K\_M & NoSys & lora-r16 & 1602 & 1982.3 & 0.183 & 0.181 & 0.494 & 0.494 & 0.004 & 0.002 \\
 L-3.2 1B & Q4\_K\_M & NoSys & lora-r8 & 1825 & 2689.0 & 0.187 & 0.183 & 0.499 & 0.501 & 0.013 & 0.008 \\
 L-3.2 1B & Q4\_K\_M & NoSys & lora-r4 & 1305 & 1567.4 & 0.176 & 0.172 & 0.496 & 0.502 & 0.008 & 0.004 \\
 L-3.2 1B & Q4\_K\_M & NoSys & origin & 1417 & 1514.2 & 0.203 & 0.202 & 0.516 & 0.515 & 0.050 & 0.016 \\
 L-3.2 3B & FP16 & RRC\_constrain & full & 2341 & 3501.9 & 0.969 & 1.000 & 0.991 & 1.000 & 0.993 & 0.991 \\
 L-3.2 3B & FP16 & RRC\_constrain & lora-r16 & 2341 & 3502.4 & 0.972 & 1.000 & 0.991 & 1.000 & 0.998 & 0.995 \\
 L-3.2 3B & FP16 & RRC\_constrain & lora-r8 & 2335 & 3533.8 & 0.966 & 1.000 & 0.989 & 1.000 & 0.994 & 0.991 \\
 L-3.2 3B & FP16 & RRC\_constrain & lora-r4 & 2331 & 3492.7 & 0.963 & 1.000 & 0.989 & 1.000 & 0.994 & 0.992 \\
 L-3.2 3B & FP16 & RRC\_constrain & origin & 369 & 360.5 & 0.369 & 0.400 & 0.549 & 0.548 & 0.328 & 0.035 \\
 L-3.2 3B & FP16 & RRC & full & 2157 & 2939.1 & 0.408 & 0.353 & 0.688 & 0.700 & 0.000 & 0.000 \\
 L-3.2 3B & FP16 & RRC & lora-r16 & 1467 & 1930.6 & 0.233 & 0.229 & 0.555 & 0.549 & 0.003 & 0.001 \\
 L-3.2 3B & FP16 & RRC & lora-r8 & 1566 & 1645.1 & 0.206 & 0.209 & 0.556 & 0.562 & 0.031 & 0.010 \\
 L-3.2 3B & FP16 & RRC & lora-r4 & 1651 & 1647.4 & 0.218 & 0.221 & 0.527 & 0.524 & 0.035 & 0.013 \\
 L-3.2 3B & FP16 & RRC & origin & 1645 & 1676.8 & 0.221 & 0.220 & 0.550 & 0.548 & 0.023 & 0.009 \\
 L-3.2 3B & FP16 & NoSys & full & 1755 & 1848.0 & 0.214 & 0.223 & 0.466 & 0.467 & 0.011 & 0.004 \\
 L-3.2 3B & FP16 & NoSys & lora-r16 & 1894 & 1906.7 & 0.196 & 0.200 & 0.524 & 0.537 & 0.034 & 0.012 \\
 L-3.2 3B & FP16 & NoSys & lora-r8 & 1962 & 2021.1 & 0.203 & 0.207 & 0.524 & 0.520 & 0.116 & 0.034 \\
 L-3.2 3B & FP16 & NoSys & lora-r4 & 1983 & 2084.9 & 0.206 & 0.204 & 0.518 & 0.518 & 0.154 & 0.091 \\
 L-3.2 3B & FP16 & NoSys & origin & 1946 & 2005.4 & 0.202 & 0.205 & 0.499 & 0.499 & 0.015 & 0.008 \\
 L-3.2 3B & Q4\_K\_M & RRC\_constrain & full & 1733 & 2658.1 & 0.954 & 1.000 & 0.987 & 1.000 & 0.973 & 0.968 \\
 L-3.2 3B & Q4\_K\_M & RRC\_constrain & lora-r16 & 1730 & 2629.4 & 0.969 & 1.000 & 0.989 & 1.000 & 0.985 & 0.984 \\
 L-3.2 3B & Q4\_K\_M & RRC\_constrain & lora-r8 & 1681 & 2628.2 & 0.914 & 0.948 & 0.982 & 0.999 & 0.671 & 0.666 \\
 L-3.2 3B & Q4\_K\_M & RRC\_constrain & lora-r4 & 1723 & 2562.1 & 0.867 & 0.941 & 0.976 & 0.999 & 0.748 & 0.743 \\
 L-3.2 3B & Q4\_K\_M & RRC\_constrain & origin & 280 & 302.4 & 0.306 & 0.286 & 0.590 & 0.601 & 0.390 & 0.046 \\
 L-3.2 3B & Q4\_K\_M & RRC & full & 2170 & 2920.6 & 0.440 & 0.400 & 0.689 & 0.698 & 0.001 & 0.000 \\
 L-3.2 3B & Q4\_K\_M & RRC & lora-r16 & 1105 & 1117.6 & 0.212 & 0.222 & 0.524 & 0.528 & 0.005 & 0.002 \\
 L-3.2 3B & Q4\_K\_M & RRC & lora-r8 & 1224 & 1264.1 & 0.211 & 0.210 & 0.571 & 0.558 & 0.052 & 0.022 \\
 L-3.2 3B & Q4\_K\_M & RRC & lora-r4 & 1192 & 1223.1 & 0.216 & 0.215 & 0.536 & 0.533 & 0.004 & 0.001 \\
 L-3.2 3B & Q4\_K\_M & RRC & origin & 1105 & 1156.1 & 0.215 & 0.212 & 0.562 & 0.557 & 0.061 & 0.026 \\
 L-3.2 3B & Q4\_K\_M & NoSys & full & 1367 & 1452.8 & 0.201 & 0.207 & 0.497 & 0.493 & 0.040 & 0.015 \\
 L-3.2 3B & Q4\_K\_M & NoSys & lora-r16 & 1328 & 1387.4 & 0.202 & 0.216 & 0.548 & 0.543 & 0.033 & 0.020 \\
 L-3.2 3B & Q4\_K\_M & NoSys & lora-r8 & 1518 & 1556.5 & 0.189 & 0.197 & 0.514 & 0.514 & 0.058 & 0.022 \\
 L-3.2 3B & Q4\_K\_M & NoSys & lora-r4 & 1491 & 1517.3 & 0.206 & 0.207 & 0.513 & 0.512 & 0.164 & 0.098 \\
 L-3.2 3B & Q4\_K\_M & NoSys & origin & 1371 & 1402.1 & 0.209 & 0.211 & 0.520 & 0.515 & 0.026 & 0.011 \\
\end{longtable}
}

\end{document}